\documentclass[aps, pra, showpacs, twocolumn, amsfonts, amsmath, amssymb, floatfix,superscriptaddress, groupedaddress]{revtex4}
\usepackage{graphicx}
\usepackage{epstopdf}
\usepackage{epsfig}
\usepackage{dcolumn}
\usepackage{bm}
\usepackage{hyperref}
\usepackage{amsmath}
\usepackage{color}

\newcommand{\be}{\begin{equation}}
\newcommand{\ee}{\end{equation}}
\renewcommand{\r}{{\bf r}}

\begin{document}


\title{Cooperative effects and disorder:\\  A scaling analysis of the spectrum of the effective atomic Hamiltonian}
\author{L. Bellando$^{1}$, A. Gero$^{2,3}$, E. Akkermans$^{2}$, and R. Kaiser}
\affiliation{Universit�{\'e} de Nice Sophia Antipolis, CNRS, Institut Non-Lin{\'e}aire de Nice, UMR 7335, F-06560 Valbonne, France \\ $^{2}$Department of Physics, Technion - Israel Institute of Technology, 32000 Haifa, Israel \\ $^{3}$Department of Education in Technology and Science, Technion - Israel Institute of Technology, 32000 Haifa, Israel }

\begin{abstract}
We study numerically the spectrum of the  non-Hermitian effective Hamiltonian that describes the dipolar interaction of a gas of $N\gg 1$ atoms with the  radiation field. We analyze the interplay between cooperative effects and disorder for both  scalar and vectorial radiation fields. We show that for dense gases, the resonance width distribution follows, both in the scalar and vectorial cases,  a power law  $P(\Gamma) \sim \Gamma^{-4/3}$ that originates from cooperative effects between more than two atoms. This power law  is different from the  $ P(\Gamma) \sim \Gamma^{-1}$ behavior, which has been considered as a  signature of Anderson localization of light in random systems. We  show that in dilute clouds, the center of the energy distribution is described by Wigner's semicircle law in the scalar and vectorial cases. For dense gases, this law is replaced in the vectorial case by the  Laplace distribution. Finally, we show that in the scalar case the degree of resonance overlap  increases as a power law of  the system size for dilute gases, but decays exponentially with the system size for dense clouds. 
\end{abstract}

\pacs{42.25.Dd,42.50.Nn,72.15.Rn}

\date{\today}

\maketitle

\section{Introduction}

Photon localization in cold atomic gases shows up as an overall decrease of photon escape rates from disordered media. The different roles played by cooperative effects, such as superradiance and subradiance \cite{dicke, gross}, and disorder \cite{anderson, gang} in $d$-dimensional atomic gases have been recently investigated \cite{PRL2, EPL, PRA2}. In two and three dimensions, by considering the photon escape rate,  it has been shown that photon localization, namely the trapping of a photon inside the gas for long periods of time,  is primarily determined by cooperative effects rather than disorder. Moreover, localization occurs as a smooth crossover between delocalized and localized photons and not as a disorder-driven phase transition as expected on the basis of Anderson localization \cite{PRL2, skipetrov, PRA2}. In one dimension, due to cooperative effects and not disorder, the single-atom limit is never reached and the photons are always localized in the gas \cite{EPL}. We note that these studies on  photon escape rates  have considered the interaction of a scalar radiation field with the atoms.

Photon escape rates from an atomic gas are determined by the time evolution of the ground-state population  associated with the reduced atomic density operator of the gas. This time evolution is governed by the spectrum of the imaginary part of the effective Hamiltonian that describes the atomic system \cite{PRL2}. Unlike previous studies mentioned above, in this paper we investigate the eigenvalues of the total effective Hamiltonian. It should be noted that for an  ensemble of more than two atoms, the imaginary parts of the eigenvalues (width of the eigenstates) of the total effective Hamiltonian do not coincide with the eigenvalues of the imaginary part of the effective Hamiltonian.
Furthermore, we consider the more realistic case where the vectorial properties of the electromagnetic wave are  taken into account and compare the results to a scalar description of the  light-matter interaction. 
By a numeric diagonalization  of the non-Hermitian effective Hamiltonian, we analyze the interplay between cooperative effects and disorder in the vectorial case and compare the findings to those of the scalar case. We will show that for dense gases, the  resonance width distribution, $P(\Gamma)$, obeys, both in the scalar and vectorial cases,  a power law  $P(\Gamma) \sim \Gamma^{-4/3}$ that originates from cooperative effects between more than two atoms. This power law is different from the known  $P(\Gamma) \sim \Gamma^{-1}$  distribution, which is interpreted  as an unambiguous signature of Anderson localization of light in random systems \cite{Orlowski}. We will also show that in dilute clouds the center of the energy distribution, $P(E)$, is described by Wigner's semicircle law in the scalar and vectorial cases. For dense gases, Wigner's semicircle law is replaced  in the vectorial case by the  Laplace distribution. In all cases, however, $P(E)$ is dominated by cooperative effects, i.e.,  it is determined by the optical thickness of the sample and not by its spatial density. Finally, we will define a scaling quantity very much in the spirit of the scaling conductance $g$ introduced originally by Thouless \cite{thouless}. The quantity $g$ we consider, measures, for the effective Hamiltonian, the degree of overlap between the modes. We will show that in the scalar case the degree of resonance overlap  increases as a power law of the system size for dilute gases, but decays exponentially with the system size for dense clouds. In the vectorial case the degree of resonance overlap always increases as a power law of  the system size for both dilute and dense gases. Those results could be interpreted as a hint for the existence of a phase transition in the scalar case.

The paper is organized as follows: we start, in Section II, by describing the model which consists of $N \gg1 $ identical two-level atoms placed at random positions in an external radiation field. Then, in Section III, the effective Hamiltonian  is introduced both in the scalar and vectorial cases, and  in Section IV its spectrum is considered in the complex plane.
Later, in Sections V and VI,  the distributions $ P(\Gamma)$ and $P(E)$ are investigated. The  effect of cooperative states of more than two atoms  is studied in Section VII and the degree of resonance overlap is investigated in Section VIII. Finally, the results are discussed in Section IX.

\section{Model}
Atoms are taken as degenerate, two-level systems  denoted by $|g
\rangle= |J_{g}=0,m_{g}=0 \rangle$ for the ground state and $|e
\rangle = |J_{e}=1,m_{e}=0,\pm1 \rangle$ for  the excited state,
where $J$ is the quantum number of the total angular momentum and $m$ is its  projection
on a quantization axis, taken as the $\hat{z}$ axis. The energy
separation between the two levels, including radiative shift,  is
$\hbar\omega_{0}$ and the natural width of the excited level is
$\hbar \Gamma_{0}$.

We consider an ensemble of $N \gg1 $ identical atoms, uniformly distributed at  random positions $\textbf{r}_{i}$  in an external radiation field. The corresponding
Hamiltonian is  $H = H_0 + V$,  with  \be H_{0}={\hbar\omega_{0}} \sum_{i=1}^{N}\sum_{m_{e}=-1}^{1}( |J_{e}m_{e}\rangle\langle
J_{e}m_{e}|)_{i}+ \sum_{\bf{k}\varepsilon} \hbar \omega_{k}
a_{\bf{k}\varepsilon}^{\dag}a_{\bf{k}\varepsilon} \label{eqc1}. \ee
The light-matter interaction term $V$,
expressed in the electric dipole approximation, is
\be V=-\sum_{i=1}^{N}
\textbf{d}_{i}\cdot\textbf{E}(\textbf{r}_{i})\label{eqc2}, \ee where
${\bf E}(\r )$ is the electric-field operator
 at position $\r$,
 \be\textbf{E}(\textbf{r})=i\sum_{\bf{k}\varepsilon}\sqrt{\frac{\hbar\omega_{k}}{2\epsilon_{0}\Omega}}(a_{\bf{k}\varepsilon}\hat
\varepsilon_{\bf{k}}e^{i\bf{k}\cdot
r}-a_{\bf{k}\varepsilon}^{\dag}\hat
\varepsilon_{\bf{k}}^{*}e^{-i\bf{k}\cdot r})\label{eq4}, \ee and
$\textbf{d}_{i}=e\textbf{r}_i$ is the electric dipole moment operator of the
$i$-th atom.
$a_{\bf{k}\varepsilon}$ and $a_{\bf{k}\varepsilon}^{\dag}$ are, respectively,
the annihilation and creation operators of a mode of the field
of wave vector $\bf{k}$, polarization $\hat \varepsilon_{\bf{k}}$,
and angular frequency $\omega_{k}=ck$.
$\Omega\ $ is a quantization volume, $\epsilon_{0}$ is the vacuum dielectric constant, and $c$ is the light speed in vacuum.

We assume that the typical speed of the atoms  is small compared to
$\Gamma_{0}/k$ but large compared to $\hbar k / \mu$ where $\mu$ is
the mass of the atom, so that it is possible to neglect the
Doppler shift and recoil effects.
In addition, retardation effects are neglected; thus each atom can
influence the others instantaneously.

\section{Effective Hamiltonian}
When tracing over the radiation degrees of freedom of the Hamiltonian $H$, the following non-Hermitian Hamiltonian is obtained for the case of a single excitation \cite{zoller, PRL2}:
\be H_{eff}=\left(\hbar\omega_{0}-i\hbar\frac{\Gamma_{0}}{2}\right)\sum_{i=1}^{N}(|e\rangle\langle e|)_{i}+\hbar\frac{\Gamma_{0}}{2}\sum_{i\neq j}V_{ij}\Delta_{i}^{+}\Delta_{j}^{-}\label{eqc13}.\ee
The operators $\Delta_i^{+} =(|e\rangle\langle g|)_i$ and $\Delta_i^{-}=(|g\rangle\langle e|)_i$ are, respectively, the atomic raising and lowering operators. 
The complex-valued random  interaction potential  $V_{ij}=\beta_{ij}-i\gamma_{ij}$  is given  by
\be \beta_{ij}={3 \over 2}  \left[ - p_{ij} {\cos k_0 r_{ij} \over k_0 r_{ij}} + q_{ij}
\left( {\cos k_0 r_{ij} \over (k_0 r_{ij})^3} + { \sin k_0 r_{ij} \over (k_0
r_{ij})^2 } \right) \right]\label{ls}\ee and
\be \gamma_{ij}={3 \over 2}  \left[  p_{ij} {\sin k_0 r_{ij} \over k_0 r_{ij}} -q_{ij}
\left( {\sin k_0 r_{ij} \over (k_0 r_{ij})^3} -  { \cos k_0 r_{ij} \over (k_0
r_{ij})^2 } \right) \right]\label{er}.\ee
For  $m_e=0$ \be
p_{ij} = \sin^{2} \theta_{ij} \,  \ \ \ \ \  q_{ij} = 1 - 3 \cos^2\theta_{ij}
\label{eqc18},\ee while for  $m_{e}=\pm1$ \be
p_{ij} = \frac{1}{2}(1+\cos^2\theta_{ij})    \,   \ \ \ \ \
q_{ij} = \frac{1}{2}(3\cos^2\theta_{ij}-1)
\label{eqc19}.\ee
Here  $r_{ij}=|\textbf{r}_{i}-\textbf{r}_{j}|$ and $\theta_{ij}=\cos^{-1}(\hat{z}\cdot\hat{r}_{ij})$. 
The effective Hamiltonian has two components. The first part is the single-atom Hamiltonian including the natural width of the excited state. The second component is the contribution of  cooperative effects between any two atoms \cite{stephen, lehmberg} when retardation is neglected \cite{milonni}. Equation (\ref{ls}) gives the cooperative level shift, while Eq. (\ref{er})  gives the cooperative correction to the single-atom spontaneous emission rate.

Averaging $V_{ij}$ over the random orientations of the pairs of atoms leads to \cite{PRL1, PRA1}
\be \beta_{ij}=-\frac{\cos k_{0}r_{ij}}{k_{0}r_{ij}}\label{s1} \ee and
\be \gamma_{ij}=\frac{\sin k_{0}r_{ij}}{k_{0}r_{ij}}\label{s2},\ee
namely, the cooperative level shift and the cooperative correction to the spontaneous emission rate in the case
where the atoms are coupled to a scalar radiation field \cite{vries}.

In order to study the complex eigenvalues $E_{n}-i\hbar\Gamma_n/2$ of $H_{eff}$, we define the complex-valued quantities $\Lambda_n$ by
\be E_{n}-i\hbar\Gamma_n/2 = \hbar\omega_0+\hbar\Gamma_0\Lambda_n. \ee
The real part of $\Lambda_{n}$ corresponds to the (properly rescaled) energy of a collective state relative to a single-atom resonance and its imaginary part is related to the decay rate of this eigenstate.
For a single atom ($N=1$), we thus have $\Lambda_1=-i/2$.

In the case of  a cooperative pair $(N=2)$, namely, two atoms separated by a distance $r=|\textbf{r}_{1}-\textbf{r}_{2}|$, the spectrum of $H_{eff}$  can be obtained explicitly. 
In the scalar case it is given by \cite{Rusek}
\be \Lambda^{(s)}_{\pm}=-\frac{1}{2}\left( i\pm\frac{e^{ik_0r}}{k_0r}\right).\label{cps} \ee
In the vectorial case, two of the eigenvalues of $H_{eff}$ are of a single multiplicity
\be \Lambda^{(v_1)}_{\pm}=-\frac{1}{2}\left[i\pm \frac{3}{2} e^{ik_0r}\left(-\frac{2i}{(k_0r)^2}+\frac{2}{(k_0r)^3}\right)\right], \label{cpv1}\ee
and the other two are  of a double multiplicity
\be \Lambda^{(v_2)}_{\pm}=-\frac{1}{2}\left[i\pm \frac{3}{2} e^{ik_0r}\left(\frac{1}{k_0r}+\frac{i}{(k_0r)^2}-\frac{1}{(k_0r)^3}\right)\right]. \label{cpv2}\ee

Let us examine the limiting cases. When the atoms are well separated ($k_0r \gg1$), then $\Lambda^{(s)}_{\pm}=\Lambda^{(v_1)}_{\pm}=\Lambda^{(v_2)}_{\pm}=-i/2$ and the single-atom spontaneous emission rate is recovered.
For $ k_0r \ll1$, namely in the Dicke regime, the spectrum is approximated by \be\Lambda^{(s)}_{\pm} \simeq -\frac{1}{2}\left[i \pm\left(i+\frac{1}{k_0r}\right)\right], \label{1}\ee \be\Lambda^{(v_1)}_{\pm} \simeq -\frac{1}{2}\left[i \pm\left(i+\frac{3}{(k_0r)^3}\right)\right]\label{2},\ee  and \be\Lambda^{(v_2)}_{\pm} \simeq -\frac{1}{2}\left[i \pm\left(i-\frac{3}{2(k_0r)^3}\right)\right].\label{3}\ee   In all cases, the imaginary part of the $\Lambda_{+}$'s ($\Lambda_{-}$'s) accounts for the superradiant (subradiant) mode.

In order to  obtain numerically the spectrum of $H_{eff}$  in ({\ref{eqc13}) beyond the case of two atoms, we consider $N \gg 1$ atoms enclosed in a cubic volume $L^{3}$. The atoms are distributed with a uniform density $\rho=N/{L}^{3}$. With the help of the resonant radiation wavelength, $\lambda=2\pi/ k_{0}$, we define the dimensionless density $\rho \lambda^3$.
Next, we introduce the Ioffe-Regel number \cite{IR}, $k_0l$, where $l$ is the photon elastic mean free path, namely $l=1/\rho\sigma$, and $\sigma$ is the average single scattering cross section. For resonant scattering, the average single-scattering cross section varies as $\lambda^2$, so that the Ioffe-Regel number can be written as  $k_0l^{(s)}=2\pi^2/ \rho \lambda^3$ in the scalar case and $k_0l^{(v)}=(2 / 3)  k_0l^{(s)}$ in the vectorial case \cite{book}.
Finally, we define the (on resonance) optical thickness, $b_0$, as the ratio between the system size $L$ and the photon elastic mean free path $l$. Using the definitions above, one obtains $b_0^{(s)}=N^{1/3} (\rho \lambda^3)^{2/3} / \pi$ and
$b_0^{(v)} = (3/2) b_0^{(s)}$.

While the Ioffe-Regel number accounts for disorder effects, cooperative effects are more accurately described by the optical thickness \cite{PRL2, skipetrov, Lin1, Lin2, Bienaime}. Therefore, we will use these two parameters in order to investigate the distinctive roles of disorder and  cooperative effects in atomic gases.

\section{Spectrum of the effective Hamiltonian}

The complex-valued spectrum of $H_{eff}$ in ({\ref{eqc13}) for optically and spatially dilute gases ($b_0\ll1$ and $\rho \lambda^3 \ll 1$) is displayed in Fig. 1 for the scalar case (top) and the vectorial case (bottom). The spiral branches (magenta curves) in the scalar case  represent the eigenvalues of cooperative pairs (\ref{cps}), while the branches in the vectorial case [green (light gray) and blue (dark gray) curves] represent the eigenvalues of cooperative pairs  (\ref{cpv1}) and (\ref{cpv2}).
Eigenvalues of  states of more than two atoms are concentrated within an ellipse on the complex plane. The parameters of the ellipse will be determined in Section VII. In dilute gases, due to the dominance  of the $1/k_0r_{ij}$ term in $V_{ij}$, there are no significant differences, except for the cooperative pairs, between the spectrum of the scalar case and the spectrum of the vectorial case.

\begin{figure}[h!]
\centering
\begin{minipage}[b]{8.5cm}
\includegraphics[width=8.5cm]{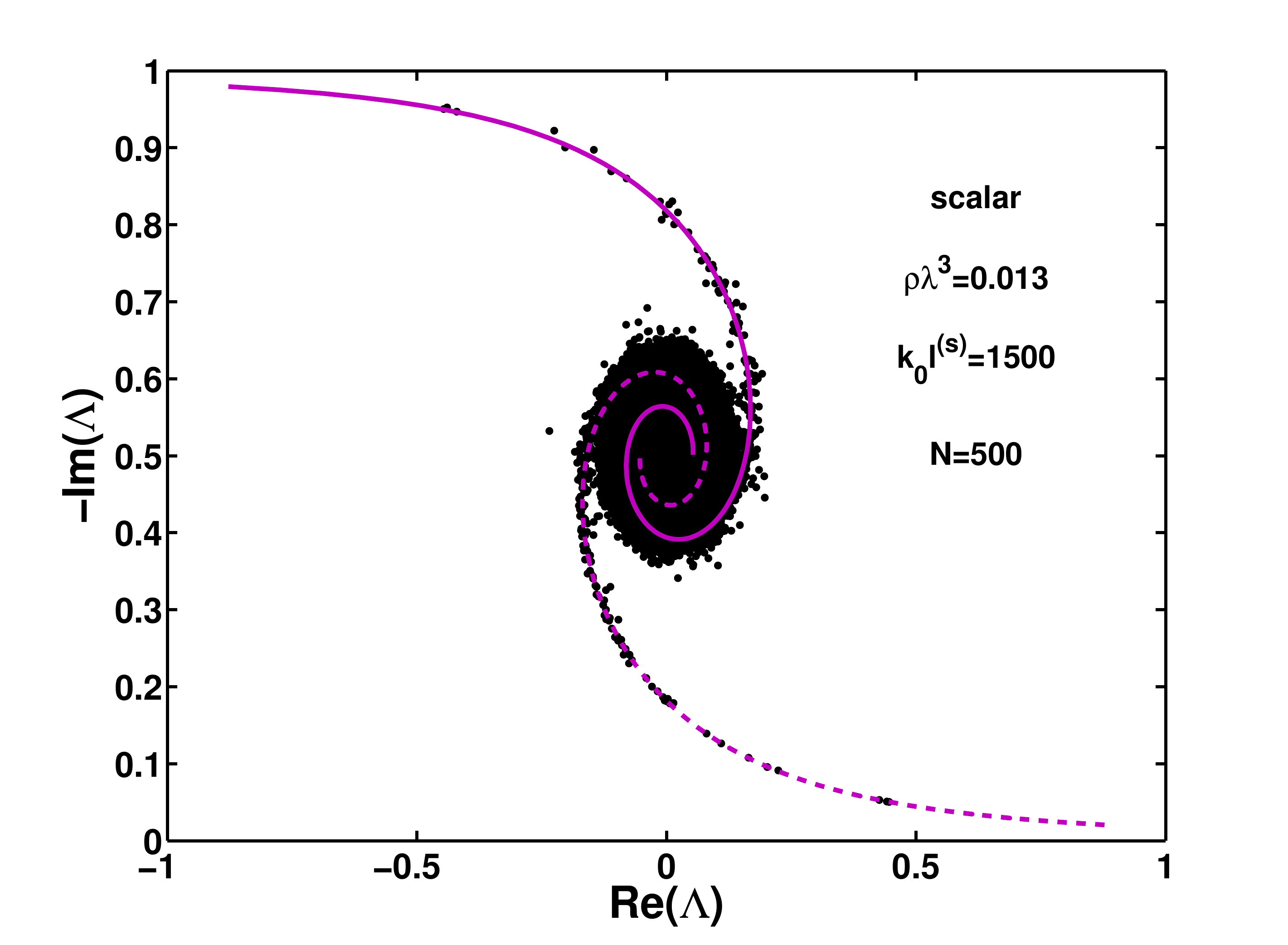}
\end{minipage}
\hspace{0.5cm}
\begin{minipage}[b]{8.5cm}
\includegraphics[width=8.5cm]{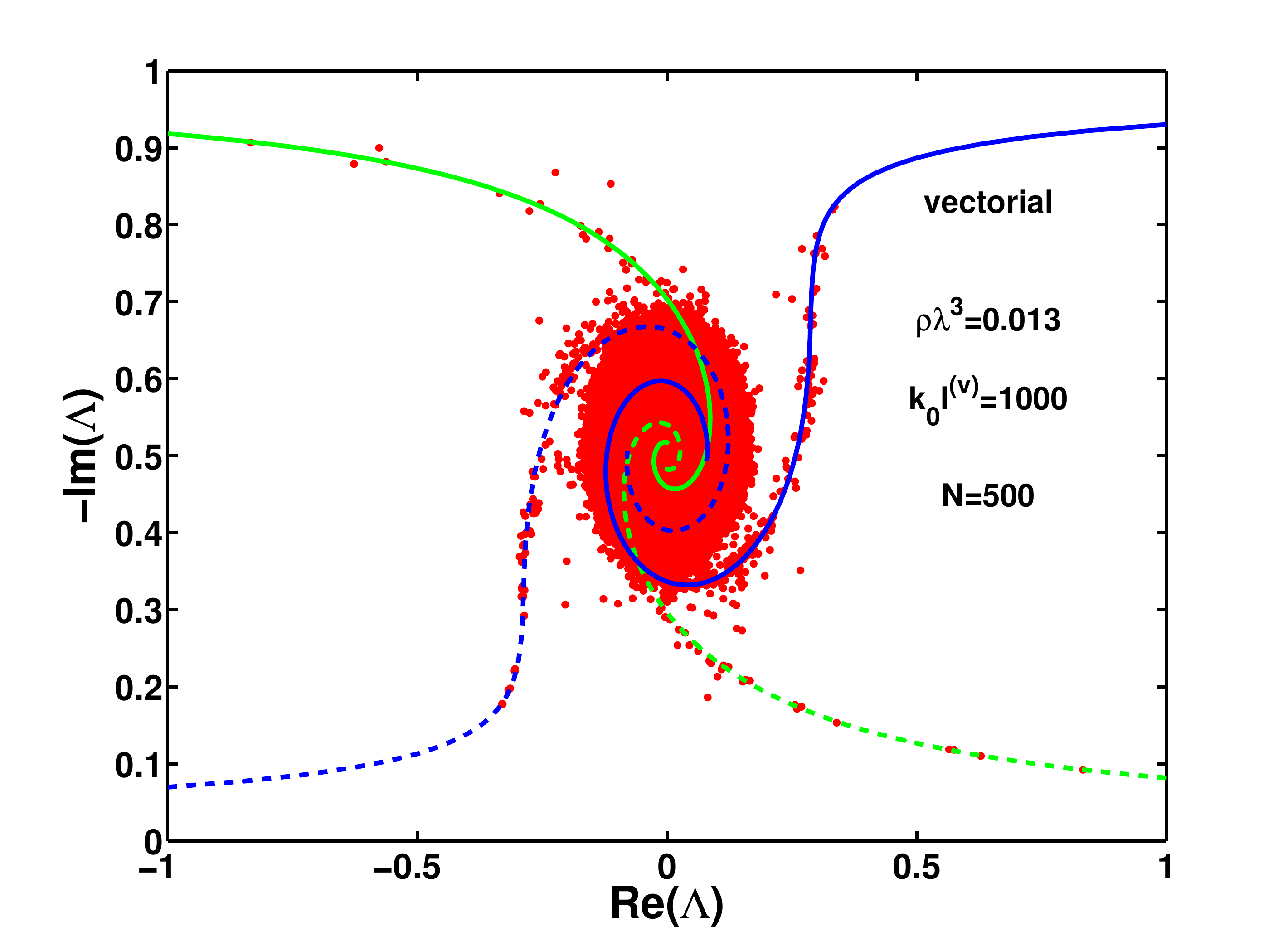}
\caption{\em (Color
   online) Complex-valued spectrum of  $H_{eff}$ (\ref{eqc13})  in the scalar case (top) and the vectorial case (bottom) for $N=500$ and $\rho \lambda^3=0.013$.  The spiral branches in the scalar case (magenta curves) represent the eigenvalues of cooperative pairs (\ref{cps}), while the branches in the vectorial case [green (light gray) and blue (dark gray) curves] represent the eigenvalues of cooperative pairs  (\ref{cpv1}) and (\ref{cpv2}).}
 \label{fig1}
\end{minipage}
\end{figure}

For optically and spatially dense gases ($b_0\gg1$ and $\rho \lambda^3 \gg 1$), however, there are remarkable  differences between the spectra obtained for the scalar and vectorial cases, as can be seen in Fig. 2.
First, we observe a disappearance of scalar superradiant pairs, while vectorial superradiant pairs persist.  Second,  there are  more vectorial subradiant pairs than scalar subradiant pairs and the former span over larger values of energy. Finally,  unlike the vectorial case, a large number of scalar subradiant states of more than two atoms  appear around the energy of $|\mbox{Re}(\Lambda)|\simeq 1$. These findings have a profound effect on the characteristics of the resonance widths, as will be discussed in Section V.

\begin{figure}[h!]
\centering
\includegraphics[width=8.5cm]{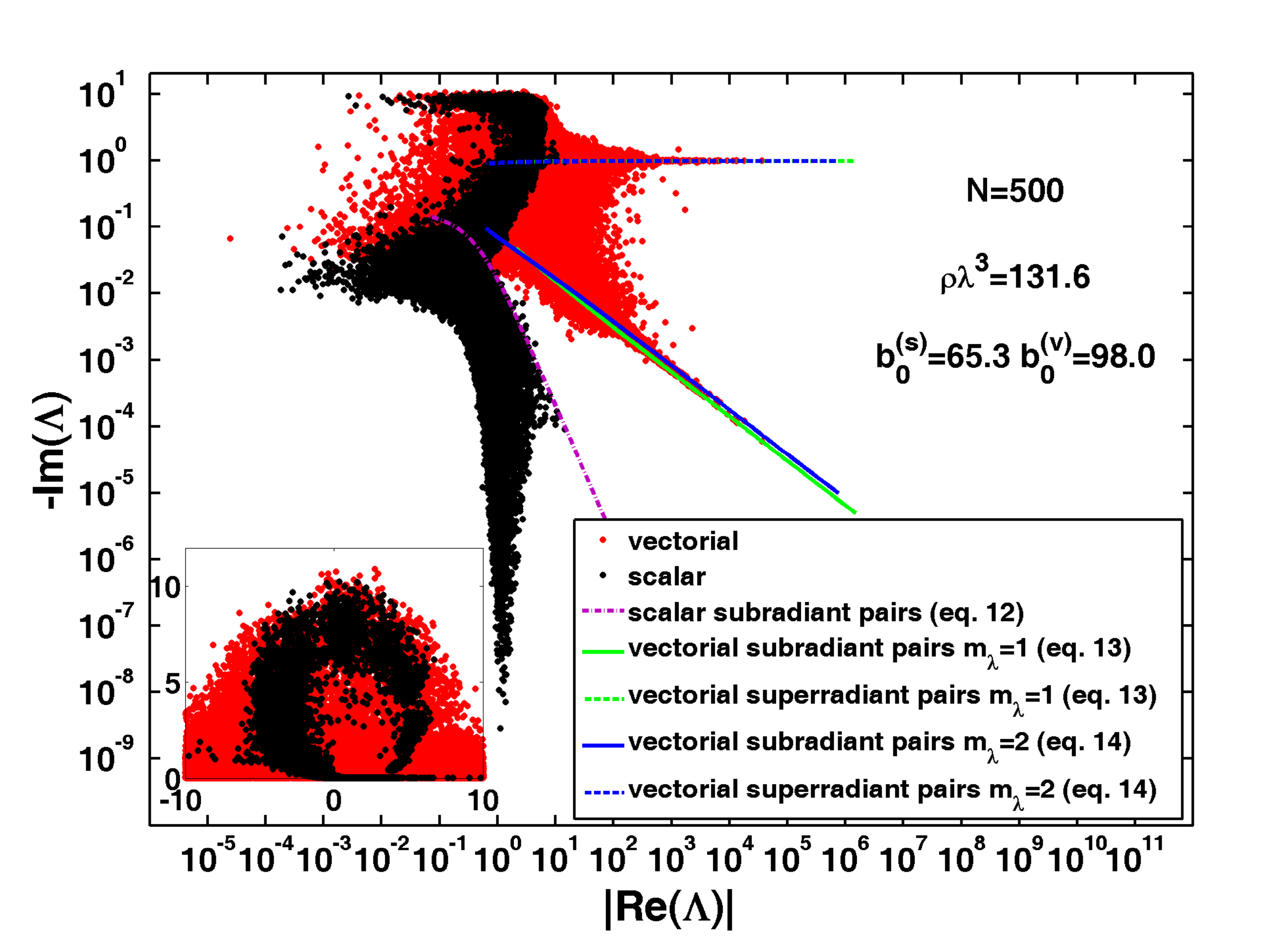}
\caption{\em (Color
   online) Complex-valued spectrum of $H_{eff}$ (\ref{eqc13}) in the scalar (black points) and vectorial [red (dark gray) points] cases for $N=500$ and $\rho \lambda^3=131.6$.  Inset: vectorial [red (dark gray) points] and scalar (black points) eigenvalues in linear scale.}
 \label{fig2}
\end{figure}

\section{Resonance width distribution}

In this section we study the resonance width distribution, $P(\Gamma)$, where $\Gamma=-2\mbox{ Im}(\Lambda)$ is a normalized resonance width (in units of $ \Gamma_0$). $P(\Gamma)$ is displayed in Fig. 3 for the scalar case (top) and the vectorial case (bottom). For dilute gases, the distribution is peaked at $\Gamma=1$ 
both in the scalar and vectorial cases, indicating the dominance of independent atoms physics. For dense clouds, when the  optical thickness is large enough, the distribution in both cases is  well described by the  power law $P(\Gamma) \sim \Gamma^{-4/3}$  rather than $P(\Gamma) \sim \Gamma^{-1}$, as suggested in \cite{Orlowski}.  We note  that this power law is obtained without taking into account the real part of the eigenvalue, and therefore it merely indicates that not all eigenstates follow a $P(\Gamma) \sim \Gamma^{-1}$ scaling \cite{Skipetrov_PC}. In Section VII, we will  further investigate the origin of this behavior.   It should be noted that regardless of the system parameters, the resonance widths  are constrained by $\langle\Gamma\rangle_i=-2\mbox{Tr} (\Lambda)/N=1 $, where  $\langle.\rangle_i$ denotes the average over the spectrum for a single realization, $i$, of atomic disorder. Additionally, as noted in  Section IV, in the vectorial case there are less long-living modes of more than two atoms compared to the scalar case.

\begin{figure}[h!]
\centering
\begin{minipage}[b]{8.5cm}
\includegraphics[width=8.5cm]{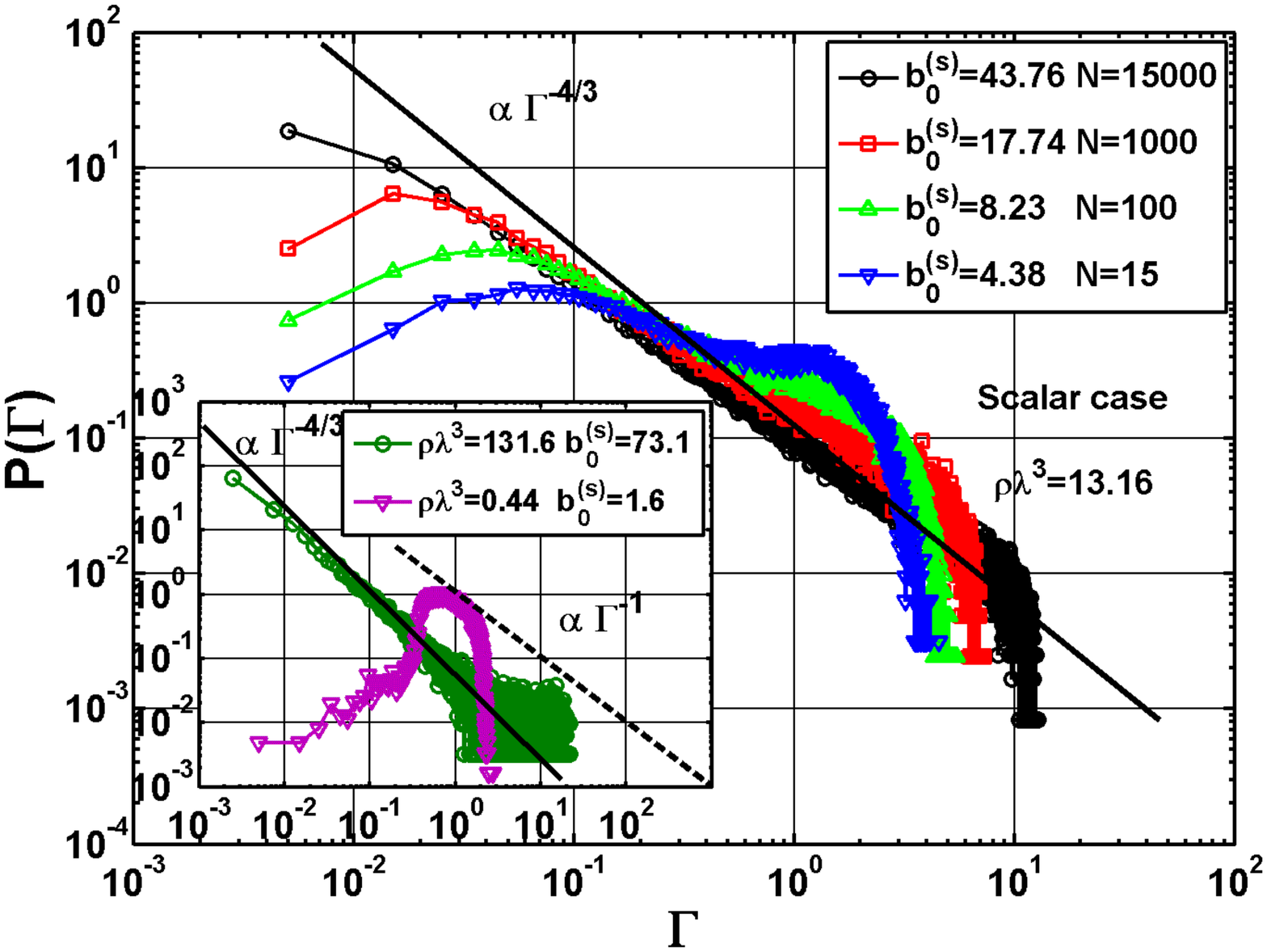}
\end{minipage}
\hspace{0.5cm}
\begin{minipage}[b]{8.5cm}
\includegraphics[width=8.5cm]{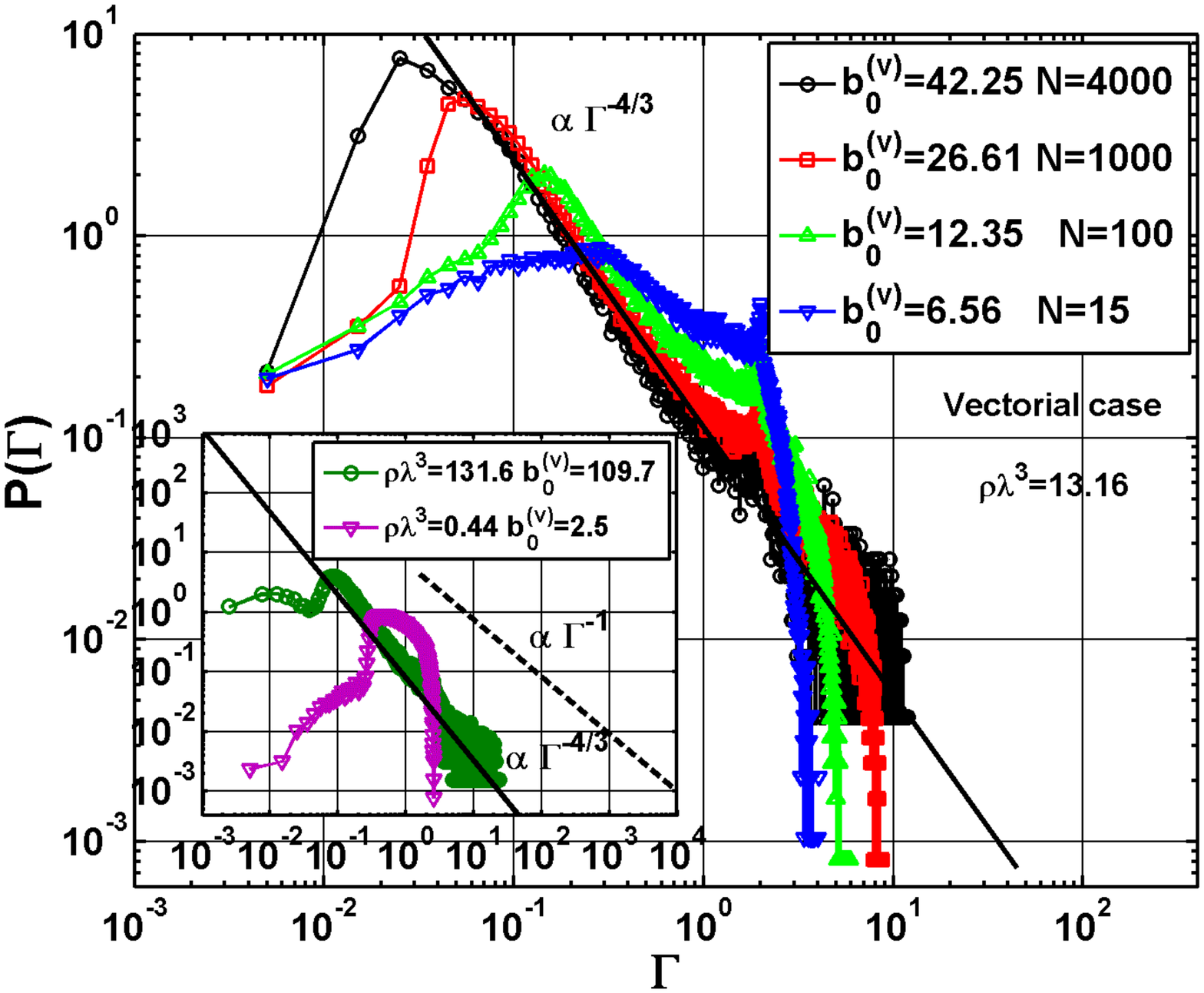}
\caption{\em (Color
   online) Resonance width distribution in the scalar case (top) and the vectorial case (bottom) for $\rho \lambda^3=13.16$.  Insets: the resonance width distribution  for $\rho \lambda^3=131.6$ and $\rho \lambda^3= 0.44$.}
 \label{fig3}
\end{minipage}
\end{figure}

Next, we study the asymptotic behavior of the  resonance widths.  For dilute gases, the configuration-averaged maximal resonance width, $\Gamma_{max}$,  and the configuration-averaged minimal resonance width, $\Gamma_{min}$, are determined by cooperative pairs as can be seen numerically in Fig. 1 and as theoretically argued  by the authors of Ref. \cite{skipetrov}. For dense clouds, as shown in Fig. 2, the effect of configurations of more than two atoms should be taken into account. Fig. 4  (top) presents  the dependence of $\Gamma_{max}$ on the optical thickness in the scalar case. Following the expression suggested in \cite{skipetrov, skipetrovb}, we use
\be \Gamma^{(s)}_{max}=\sqrt{\frac{b_0^{(s)}}{A}+\bigg(\frac{b_0^{(s)}}{B}\bigg)^2}+\frac{b_0^{(s)}}{C}+1, \label{maxs}\ee
where $A$, $B$ and $C$ are free fitting parameters, and obtain $A=1.70$, $B=8.50$ and $C=8.90$. 
The theoretical limits predicted by the Marchenko-Pastur law \cite{skipetrov}, namely  $\Gamma^{(s)}_{max}\propto\sqrt{b_{0}^{(s)}}$ for low optical thickness $ b_{0}^{(s)}$ and $\Gamma^{(s)}_{max}\propto b_{0}^{(s)}$ for  high optical thickness, can be recovered from (\ref{maxs}).
We note that in all regimes which we were able to explore, $\Gamma^{(s)}_{max}$ depends solely on $b_0^{(s)}$, i.e., it is dominated by cooperative effects without a spatial density dependence.
In the vectorial case, shown in Fig. 4 (bottom), $\Gamma_{max}$  depends  both on the optical thickness and the Ioffe-Regel number and is empirically given by
\begin{eqnarray}
\Gamma^{(v)}_{max}=\sqrt{\frac{b_0^{(v)}-2/k_0l^{(v)}}{A'}+\bigg(\frac{b_0^{(v)}-2/k_0l^{(v)}}{B'}\bigg)^2} \nonumber \\
+\frac{b_0^{(v)}-2/k_0l^{(v)}}{C'}+1, \label{max}
\end{eqnarray} where $A'=1.50$, $B'=15.25$ and $C'=8.48$. 
In order to  obtain  (\ref{max}), we have used  (\ref{maxs}) with slight modifications due to disorder (i.e., density) effects. From Eq. (\ref{max}) one can see that $\Gamma^{(v)}_{max}$  is dominated  by cooperative effects, depending on the optical thickness, and slightly corrected by disorder effects, depending on the spatial density of the cloud.

The maximal resonance widths have been studied in \cite{skipetrov, skipetrovb, goetschy}.  We note, however,  that in the regimes explored in this paper,  expressions 
(\ref{maxs}) and (\ref{max}) provide a simple and adequate description of these quantities.

\begin{figure}[h!]
\centering
\begin{minipage}[b]{8.5cm}
\includegraphics[width=8.5cm]{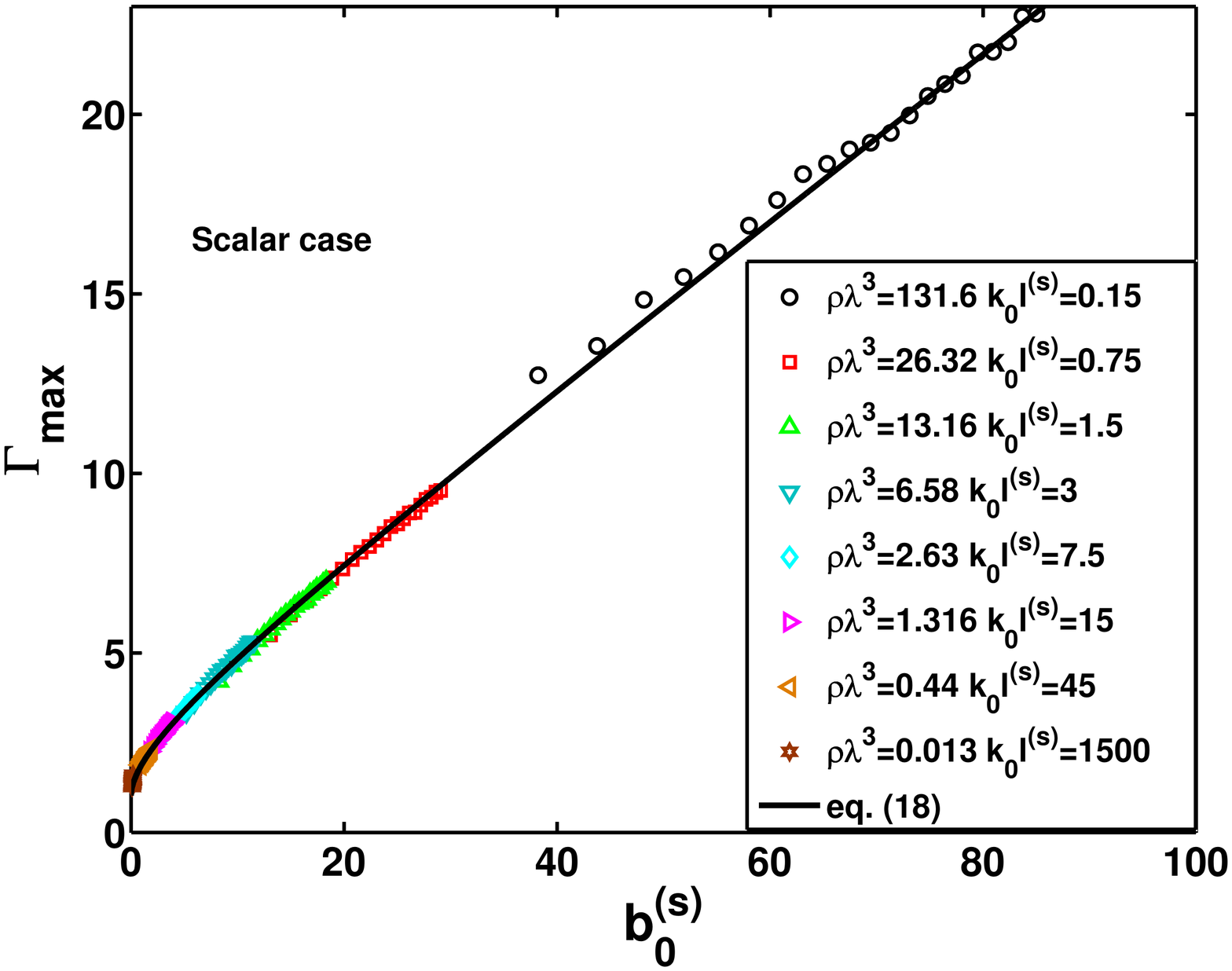}
\end{minipage}
\hspace{0.5cm}
\begin{minipage}[b]{8.5cm}
\includegraphics[width=8.5cm]{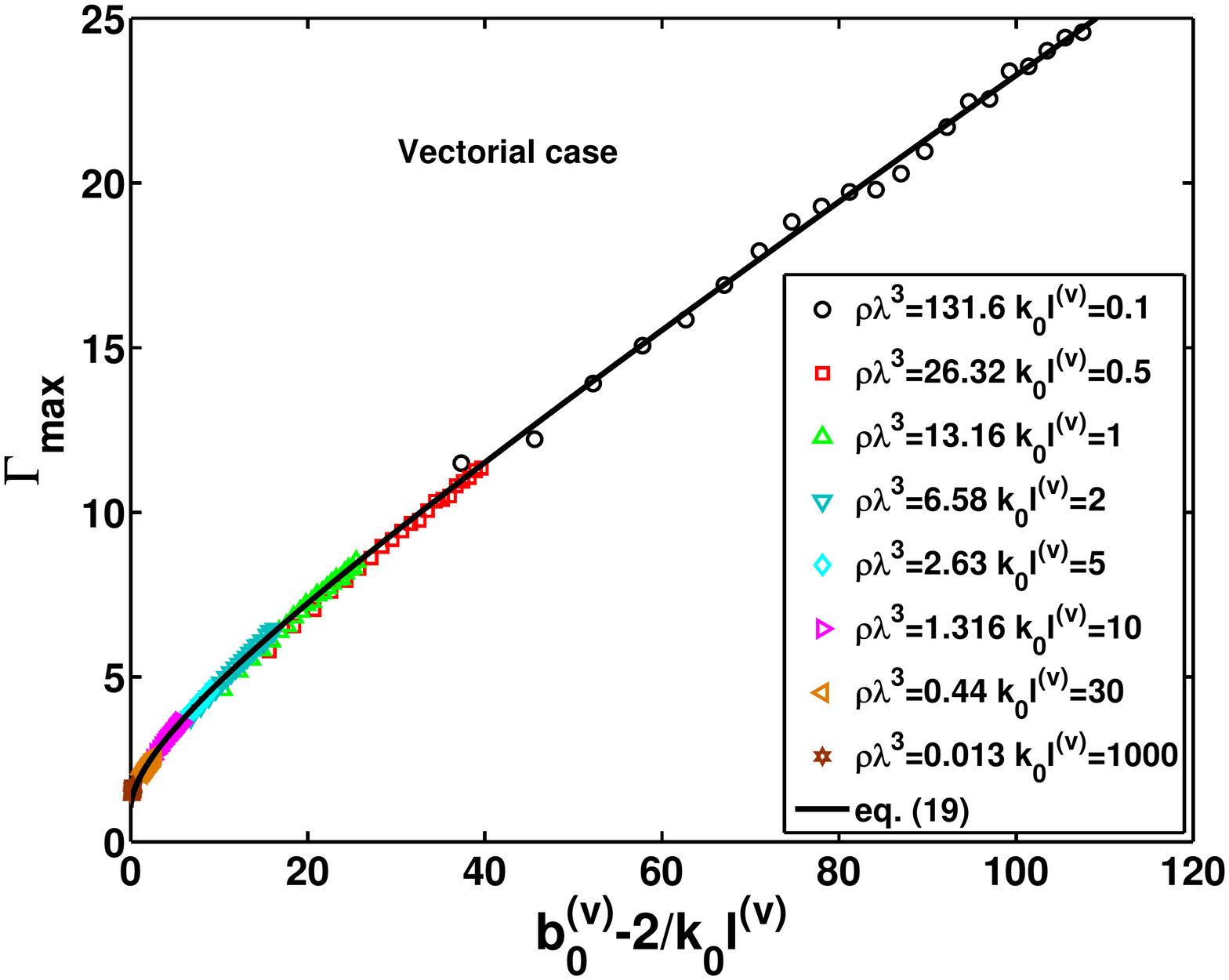}
\caption{\em (Color
   online) Behavior of $\Gamma_{max}$  in the scalar case (top)  and in the vectorial case (bottom). The solid line  is given respectively by  (\ref{maxs}) and (\ref{max}) in the scalar and vectorial cases.}
 \label{fig4}
\end{minipage}
\end{figure}

The value of $\Gamma_{min}$ in  dilute gases is determined by subradiant pairs and is given by \cite{skipetrov, goetschy}
\be \Gamma_{min} \simeq a(\rho\lambda^3 N)^{-2/3}\label{min},\ee with $a\simeq 2.30$.
As can be seen on Fig. 5 (top), Eq. (\ref{min}) holds in the scalar case  for low densities, but breaks down for dense gases. In the vectorial case (bottom), however, Eq. (\ref{min}) holds even for high densities. This difference stems from the relatively low number of scalar subradiant pairs in dense atomic clouds, as discussed in Section IV.

\begin{figure}[h!]
\centering
\begin{minipage}[b]{8.5cm}
\includegraphics[width=8.5cm]{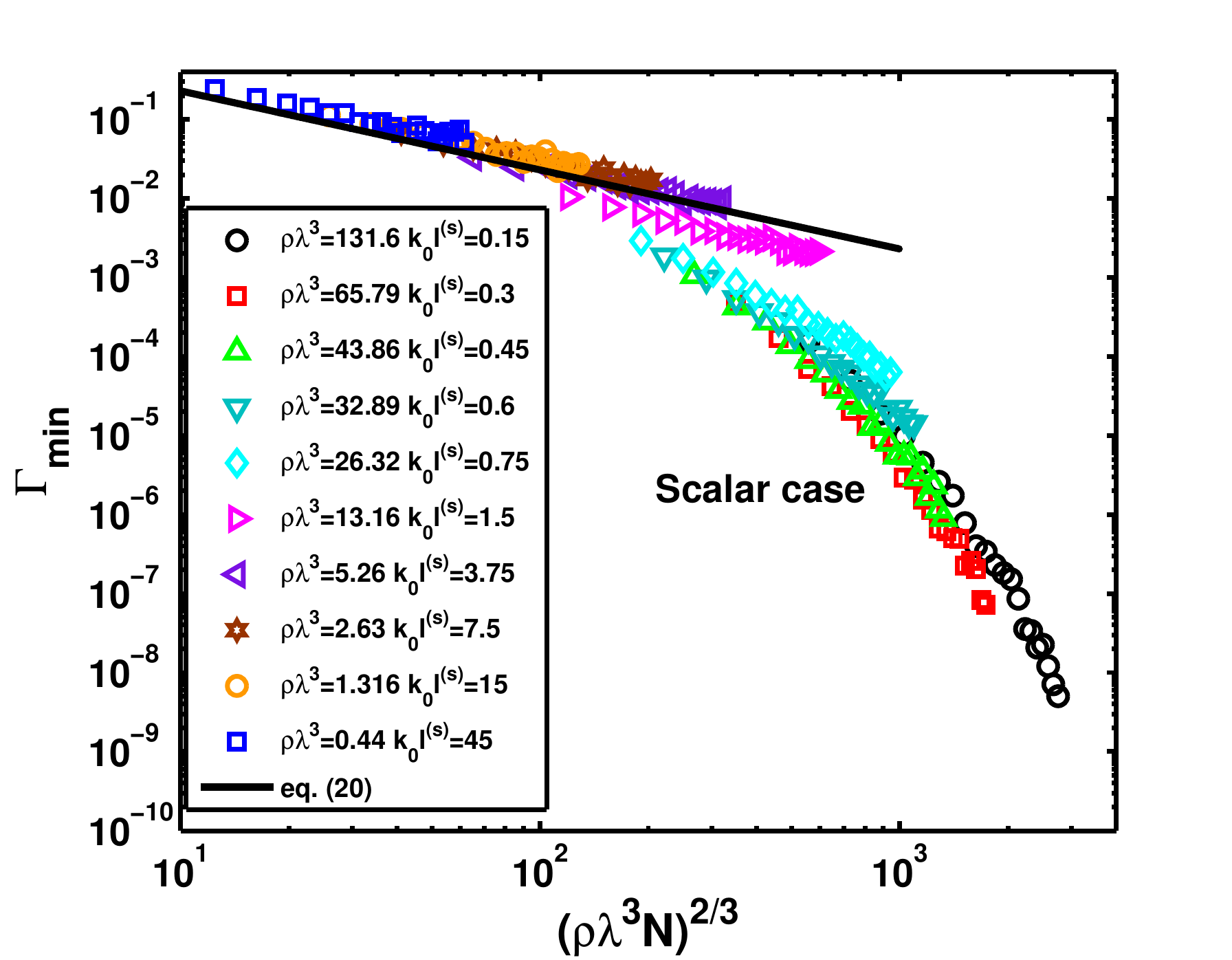}
\end{minipage}
\hspace{0.5cm}
\begin{minipage}[b]{8.5cm}
\includegraphics[width=8.5cm]{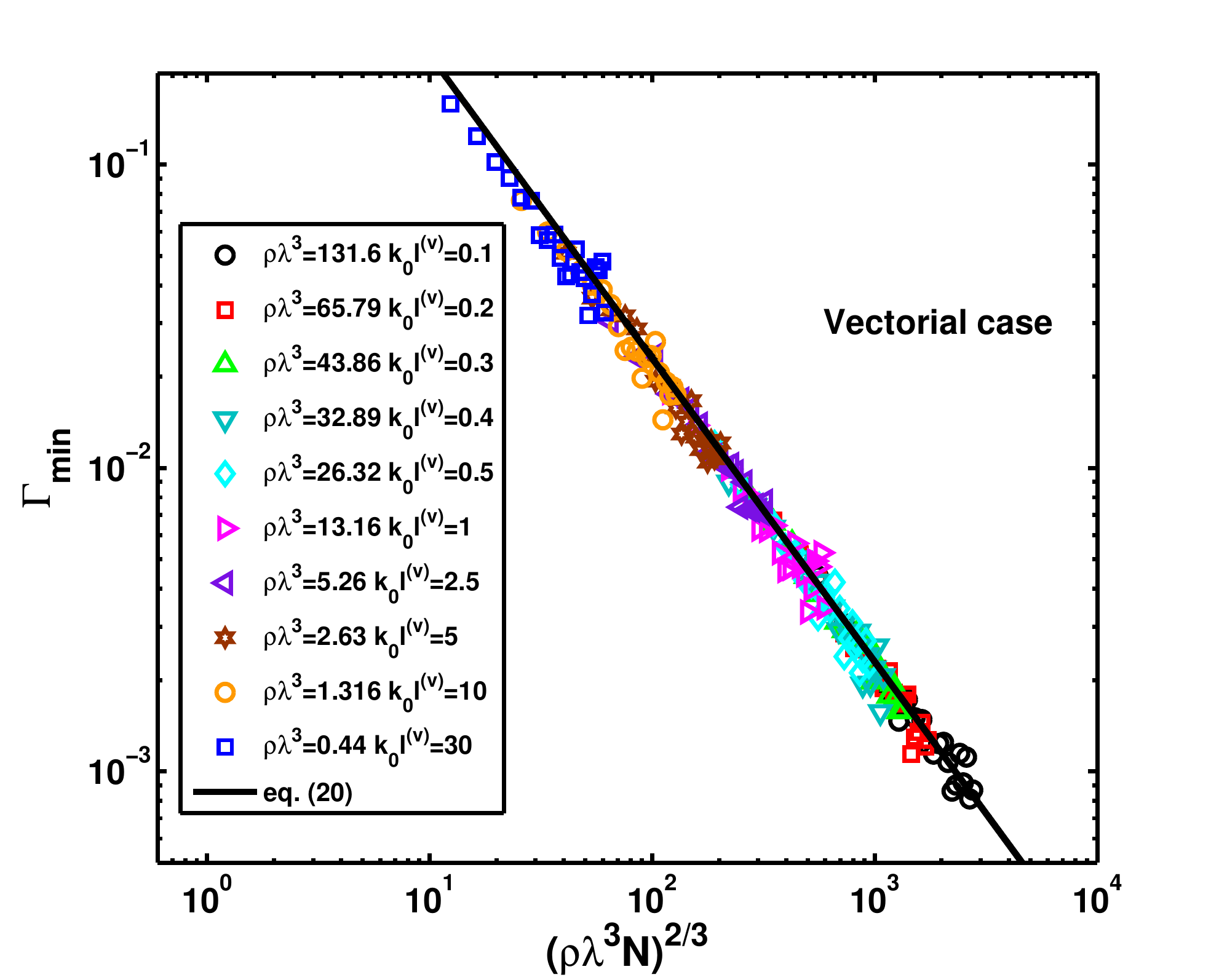}\caption{\em (Color
   online) Behavior of $\Gamma_{min}$  in the scalar case (top)  and in the the vectorial case (bottom). The solid line in both cases  is given by (\ref{min}).}
 \label{fig5}
\end{minipage}
\end{figure}

\section{Energy distribution}

Next, we study the energy distribution, $P(E)$, where $E=\mbox{ Re}(\Lambda)$ is a normalized energy (in units of $\hbar \Gamma_0$ and shifted by the atomic transition energy $\hbar \omega_0$). In dilute clouds, according to Fig. 1, cooperative pairs dominate  $P(E)$  for high values of $|E|$, both in the scalar and vectorial cases. For dense gases, as shown in Fig. 2,  due to the disappearance of scalar superradiant pairs, we expect that  for $|E| \gg 1$, $P(E)$ will be dominated by cooperative pairs  mainly in the vectorial case. In order to obtain the energy distribution of cooperative pairs, $P_{pairs}(E)$, in the limit $|E| \gg 1$, we use the relation $P_{pairs}(E)dE=P_{pairs}(r)dr$, where $P_{pairs}(r)dr=4\pi r^2 dr/ L^3$ is the  probability to find two atoms separated by a distance $r$ in the volume $L^3$ \cite{skipetrov}. We use the real part of  (\ref{1})-(\ref{3})  to calculate $dE/dr$ for $|E| \gg 1$ and find that the energy distribution of cooperative pairs in the scalar case is
\be P^{(s)}_{pairs}(E)\propto\ E^{-4}, \label {2s}\ee
and in the vectorial case is given by
\be P^{(v)}_{pairs}(E)\propto E^{-2}. \label {2v}\ee
These power laws  are indeed observed in Fig. 6, where the  energy distribution of $N=2$ atoms is calculated numerically both in the scalar case (top) and the vectorial case (bottom).

\begin{figure}[h!]
\centering
\begin{minipage}[b]{8.5cm}
\includegraphics[width=8.5cm]{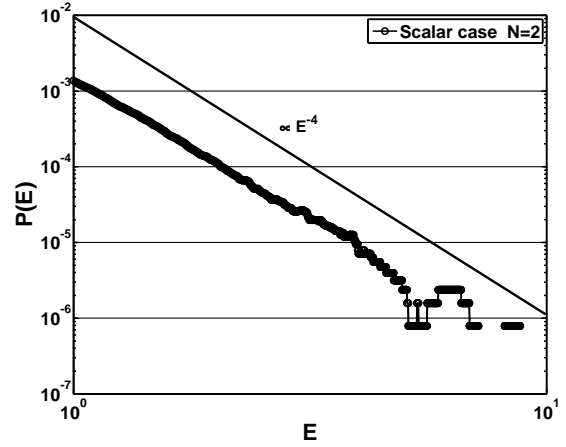}
\end{minipage}
\hspace{0.5cm}
\begin{minipage}[b]{8cm}
\includegraphics[width=8.5cm]{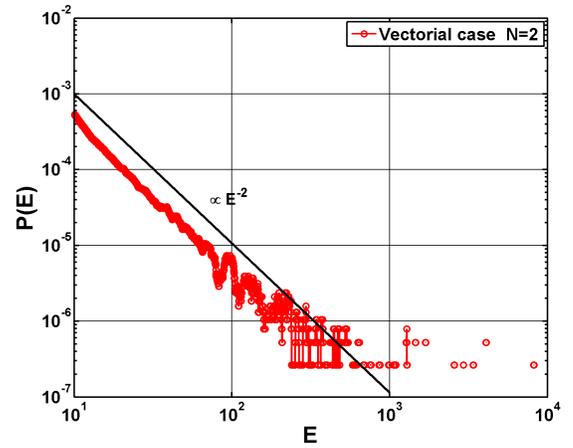} \caption{\em (Color
   online)  Energy distribution P(E) of $N=2$ atoms  in the scalar case (top)  and in  the vectorial case (bottom). The solid line  is given respectively by  (\ref{2s}) and (\ref{2v}) in the scalar and vectorial cases.}
 \label{fig6}
\end{minipage}
\end{figure}

Configurations related to more than two atoms dominate $P(E)$ for relatively low values of $|E|$. Their contribution can be described, for dilute gases, by the Wigner's semicircle law \cite{skipetrov}. Thus the total energy distribution in dilute clouds is
\be P(E)=\frac{2}{\pi}\frac{\sqrt{\alpha-E^2}}{\alpha}+P_{pairs}(E),\label {Ed}\ee
with $\alpha\simeq 0.06\ b_0^{(s)}$ ($\alpha\simeq0.10 \ b_0^{(v)}$)  in the scalar (vectorial) case.

For dense clouds, we find in the vectorial case that Wigner's semicircle law is replaced by the Laplace distribution
\be P^{(v)}(E)=\frac{e^{-\alpha'|E|}}{2\alpha'}+P^{(v)}_{pairs}(E), \label {Ev}\ee
with $\alpha'\simeq b_0^{(v)}$.
In the scalar case, however, $P(E)$ is described by the sum of the  Laplace distribution, Wigner's semicircle law, and the energy distribution of the pairs,
\be P^{(s)}(E)=\frac{e^{-\alpha'|E|}}{2\alpha'}+\frac{2}{\pi}\frac{\sqrt{\alpha-E^2}}{\alpha}+P^{(s)}_{pairs}(E).\label {Es}\ee

Since in all cases, $P(E)$ is determined solely by the optical thickness, the energy distribution is dominated by cooperative effects.
The energy distribution is shown in Fig. 7, both for the scalar case (top) and the vectorial case (bottom). By inspecting the insets, it is clear that for $|E| \gg 1$, the contribution of cooperative pairs  is indeed weaker in the scalar case compared to those in the vectorial case. We will reexamine the contribution of the pairs in the next section.

\begin{figure}[h!]
\centering
\begin{minipage}[b]{8.5cm}
\includegraphics[width=8.5cm]{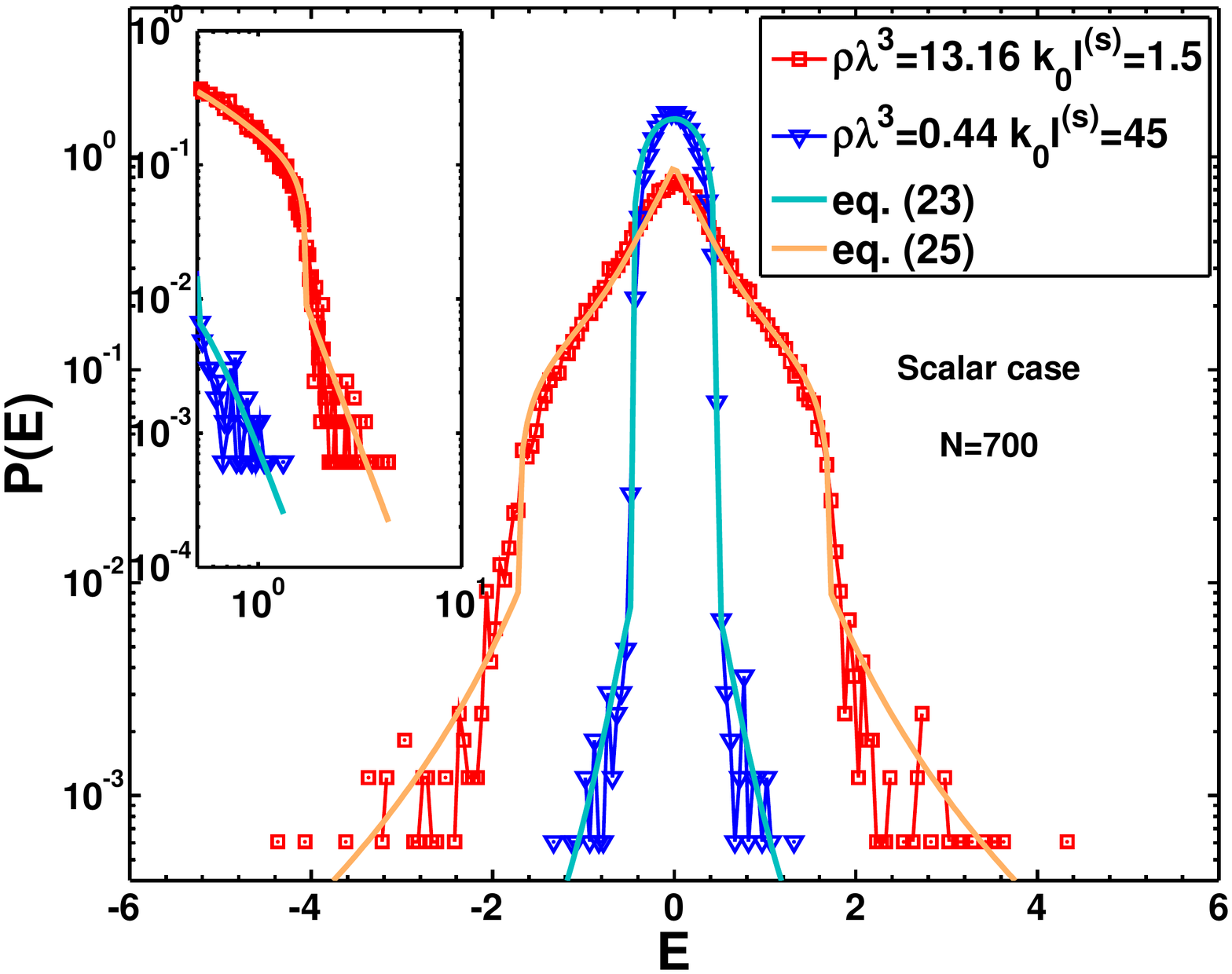}
\end{minipage}
\hspace{0.5cm}
\begin{minipage}[b]{8.5cm}
\includegraphics[width=8.5cm]{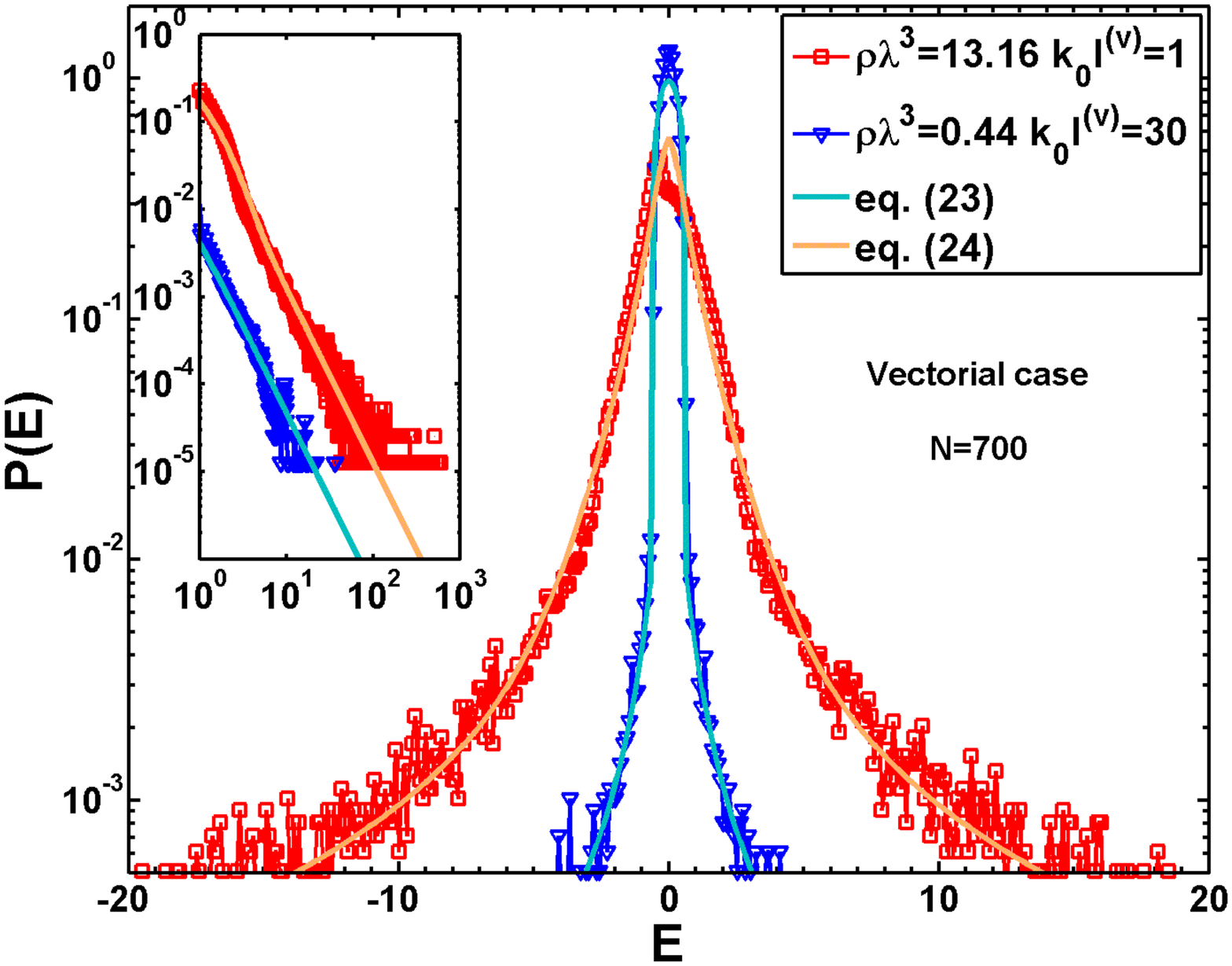} \caption{\em (Color
   online)  Energy distribution of $N=700$ atoms  in the scalar case (top) and in the the vectorial case (bottom). Dilute gases are described by  (\ref{Ed}) in both cases. Dense clouds are described by  (\ref{Es}) in the scalar case and by (\ref{Ev}) in the vectorial case.}
 \label{fig7}
\end{minipage}
\end{figure}

\section{Exclusion of cooperative pairs}

In order to disentangle between the effect of cooperative pairs and the effect of cooperative states of more than two atoms, we now exclude cooperative pairs and recalculate numerically $P(\Gamma)$ and $P(E)$.
To that purpose we characterize the ellipse on the complex plane that contains  eigenvalues related to cooperative  states of more than two atoms. The procedure described below applies to both scalar and vectorial cases, unless indicated otherwise.

We define the major axis of the ellipse as
\be \Gamma_{axe}=\frac{\Gamma_{max}+\Gamma_{fre}}{2},\ee
where the maximal resonance width is given in (\ref{maxs}) or (\ref{max}), and the most frequent resonance width is given empirically by \be \Gamma^{(v)}_{fre}=\frac{1}{b_{0}^{(s)}+1}\ee for the scalar case and
\be \Gamma^{(v)}_{fre}=\frac{1}{2b_{0}^{(v)}+1}\left(1+\frac{1}{k_0l^{(v)}}\right)\ee for the vectorial case.
The minor axis is given empirically by
\be E_{axe}=\sqrt{\frac{b_0}{D'}+\left(\frac{b_0}{D'}\right)^2},\ee
with $D'=5.50$.
All eigenvalues located inside the ellipse defined by 
\be \left (\frac{\Gamma}{\Gamma_{axe}}\right)^2+\left(\frac{E}{E_{axe}}\right)^2=1,\label{ellipse}\ee 
are related to  configurations of more than two atoms.

 In the vectorial case,  eigenvalues  located outside the domain defined by (\ref{ellipse}) are indeed mainly related to cooperative pairs. In the scalar case, however,  applying this  selection rule leads to the exclusion of the long-living modes  around $|\mbox{Re}(\Lambda)|\simeq 1$, discussed in Section IV. Thus an additional empirical criterion is used  in both cases, according to which  eigenvalues located outside a region in the complex plane whose parameters are given below are kept as well. This region is centered along curve (\ref{cps}) in the scalar case and curves (\ref{cpv1}) and (\ref{cpv2}) in the vectorial case. Its widths are $\epsilon_{sup}=1$ for superradiant pairs and $\epsilon_{sub}=\Gamma_{min}/2$  for  subradiant pairs, where $\Gamma_{min}$ is given by (\ref{min}).
 An example of such a  procedure is shown in Fig. 8  for the scalar case in the dilute (top) and dense (bottom) limits.

\begin{figure}[h!]
\centering
\begin{minipage}[b]{8.5cm}
\includegraphics[width=8.5cm]{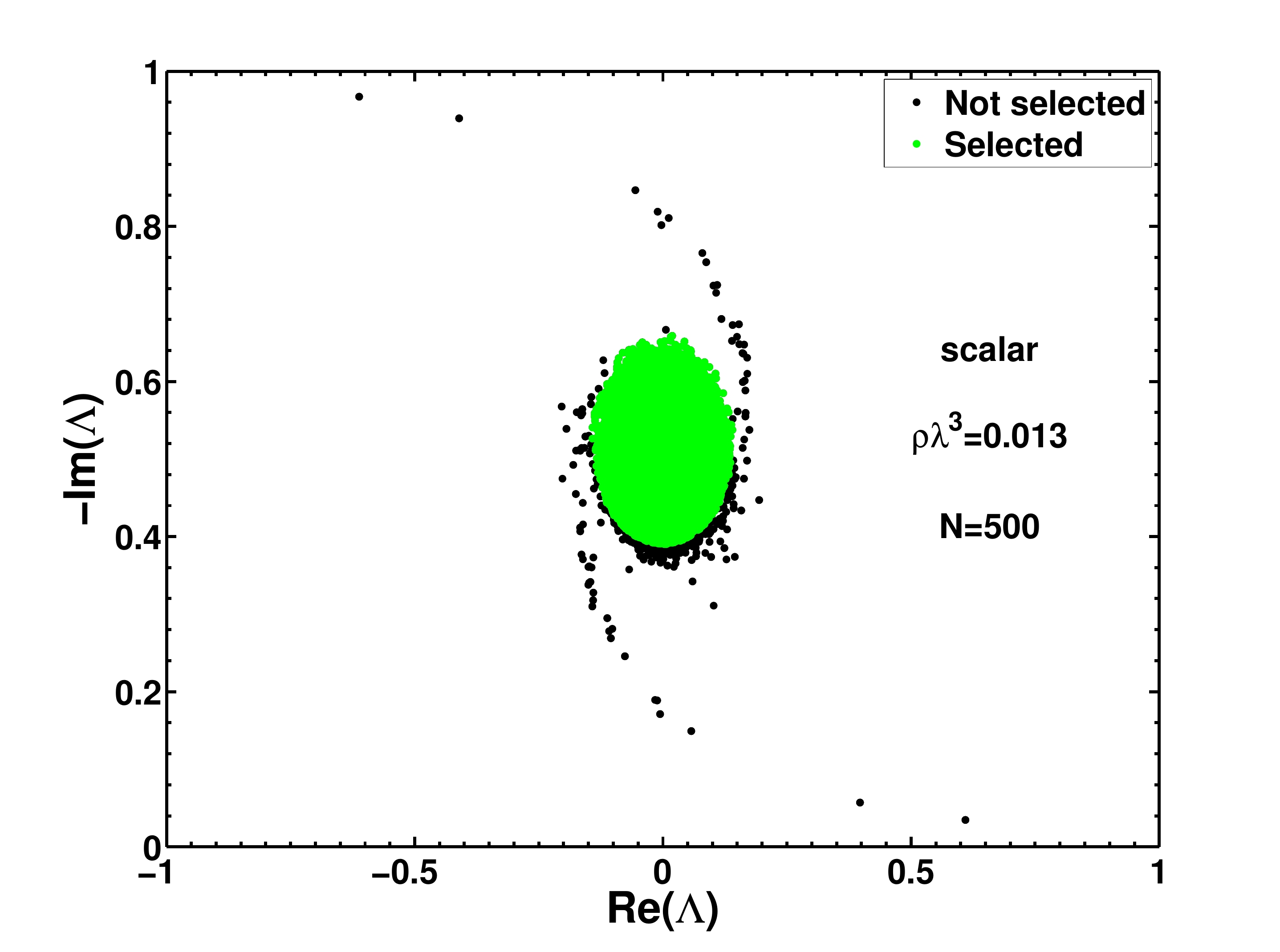}
\end{minipage}
\hspace{0.5cm}
\begin{minipage}[b]{8.5cm}
\includegraphics[width=8.5cm]{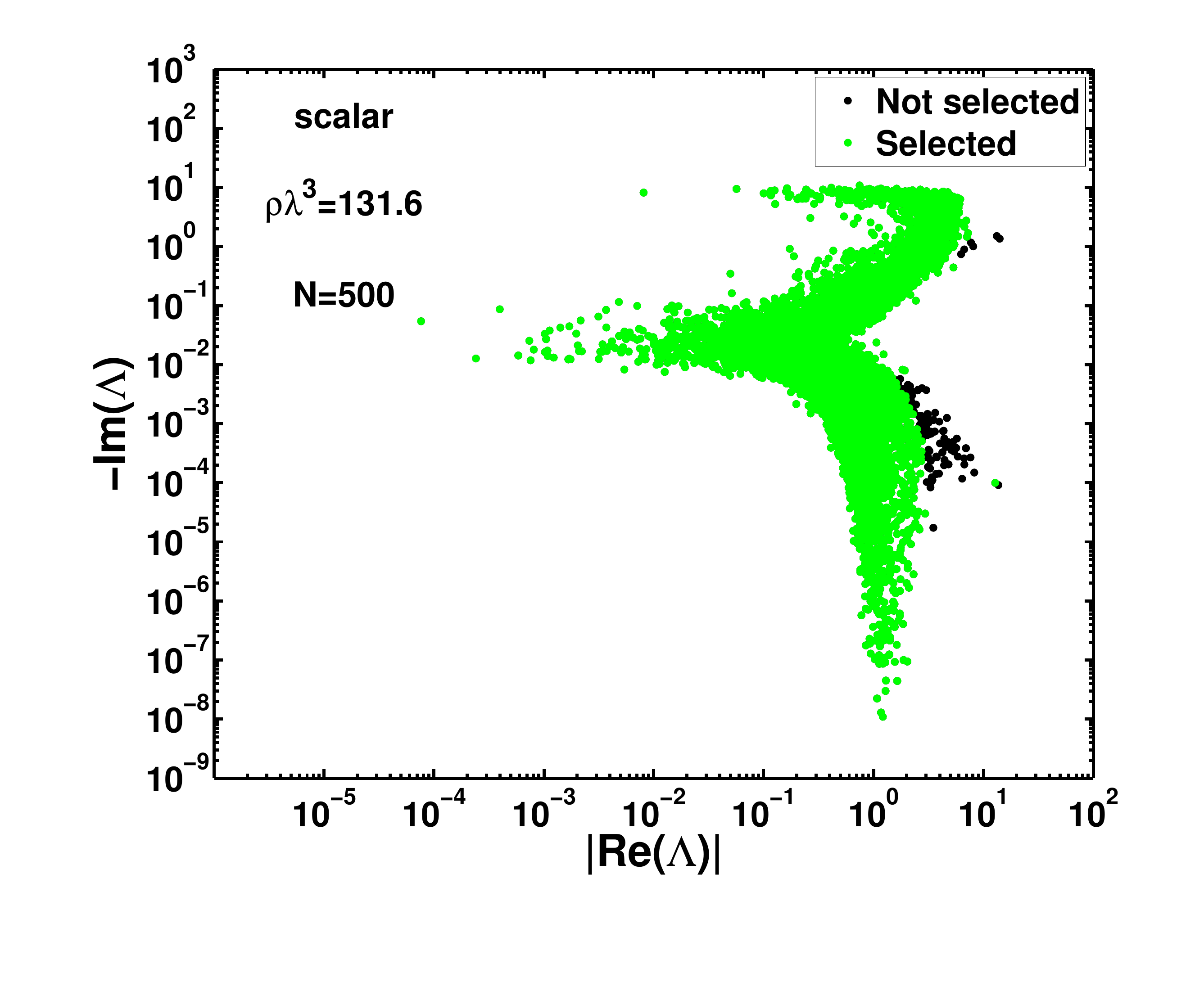}\caption{\em (Color
   online) Complex-valued spectrum of $H_{eff}$ in (\ref{eqc13}) in the scalar case  for $N=500$ and $\rho \lambda^3=0.013$  (top) and $\rho \lambda^3=131.6$ (bottom). Eigenvalues in green (light gray) (marked "Selected") are, under our assumption, related to configurations of more than two atoms. Eigenvalues in black (marked "Not selected") are related to cooperative pairs.}
 \label{fig8}
\end{minipage}
\end{figure}

\begin{figure}[h!]
\centering
\begin{minipage}[b]{8.5cm}
\includegraphics[width=8.5cm]{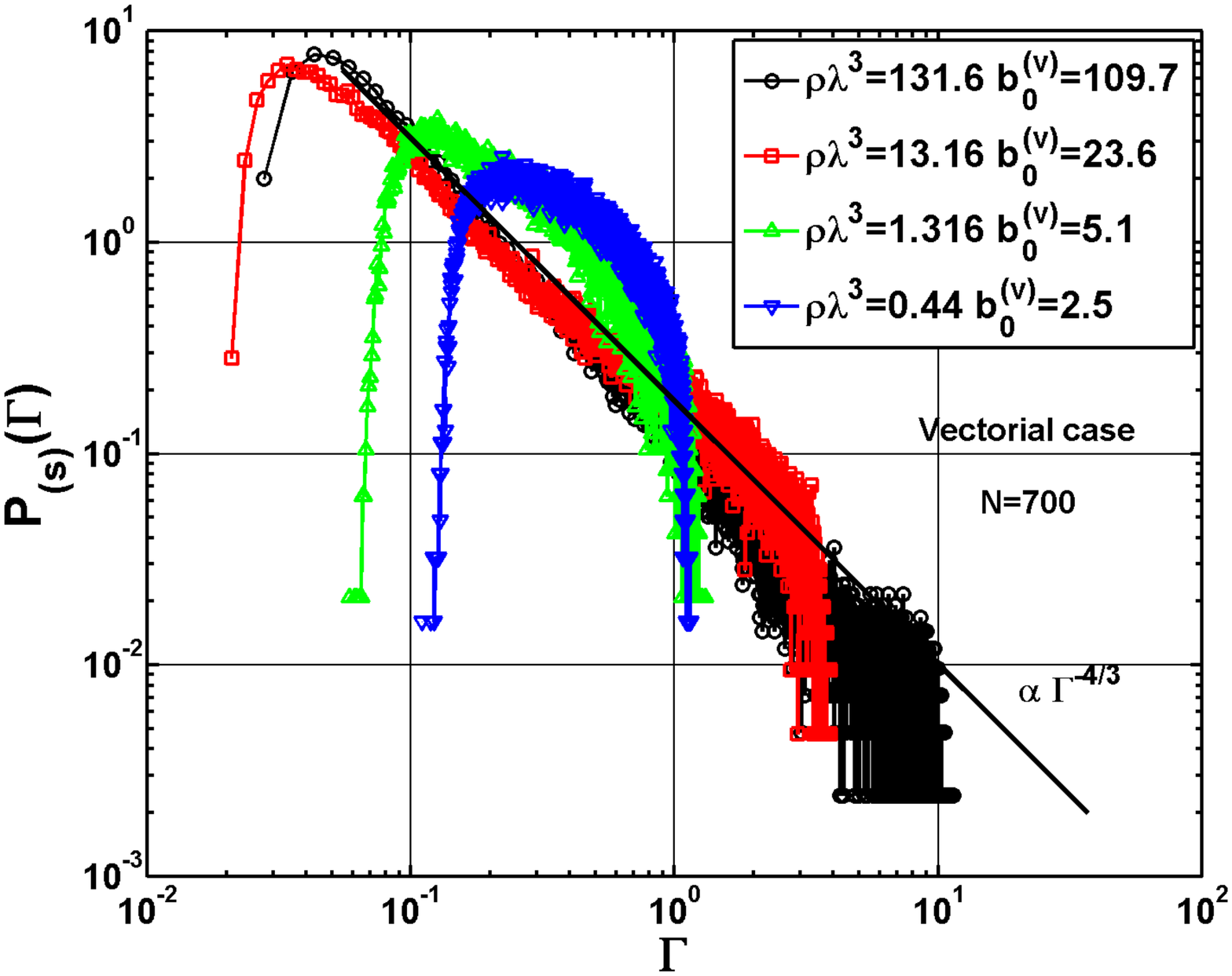}
\end{minipage}
\hspace{0.5cm}
\begin{minipage}[b]{8.5cm}
\includegraphics[width=8.5cm]{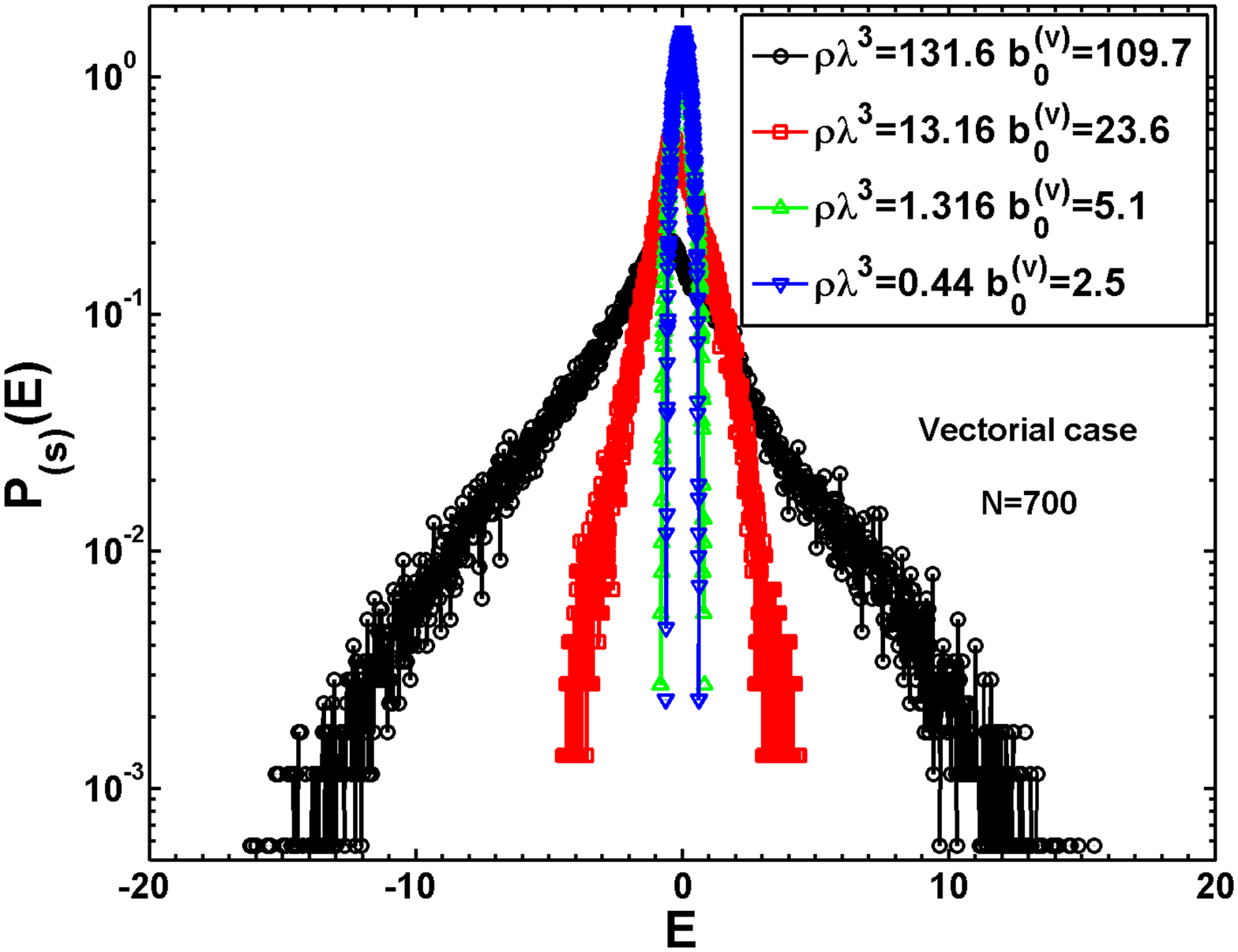} \caption{\em  (Color
   online) Vectorial case where cooperative pairs are excluded. Top: resonance width distribution. Bottom: energy distribution.}
\label{fig9}
\end{minipage}
\end{figure}


Figure 9 (top) shows the resonance width distribution in the vectorial case where cooperative pairs are excluded. It is clear that the power law $P(\Gamma) \sim \Gamma^{-4/3}$, discussed in Section  V, holds, indicating that it does not stem from cooperative pairs. Figure 9 (bottom) describes the energy distribution in the vectorial case where pairs are excluded. By comparing it to Fig. 7, the disappearance of the  $ E^{-2}$ behavior, related to cooperative pairs,  is obvious.

\section{Resonance overlap}

In this section we further analyze the statistics of the eigenvalues of $H_{eff}$ given in  (\ref{eqc13}) in order to find 
a scaling parameter which would monitor the phase transition (or lack thereof)  between localized and extended states. We will follow ideas introduced by Thouless \cite{thouless}, who showed that under specific circumstances, the inverse of the electronic dimensionless conductance $g$ (in units of $e^2/h$) can be understood as the ratio between the average level spacing between neighboring disordered energies and their widths induced by the opening of the system.

Following \cite{Genack, goetschy}, we define the degree of resonance overlap, a quantity formally analogous to the Thouless conductance,  by
\be g=\left\langle\frac{1}{\langle2/\Gamma\rangle_i\langle\Delta E\rangle_i}\right\rangle,\label{g} \ee where $\langle\Delta E\rangle_i$ is the nearest-neighbor average level spacing and $\langle2/\Gamma\rangle_i$ is the average of the inverse modal leakage rate.
Here $\langle.\rangle_i$ denotes the average over the spectrum for a single realization, $i$, of atomic disorder  and  $\langle.\rangle$ denotes the average over all configurations.
We note that this definition of the degree of resonance overlap differs from  the ratio of the average level width $\langle\Gamma\rangle_i$ to the average level spacing $\langle\Delta E\rangle_i$, used in \cite{thouless} to characterize electronic transport. The latter may not be  relevant here  since the resonance width $\Gamma$ are constrained by $\langle\Gamma\rangle_i=1 $, as mentioned in Section V.
Definition (\ref{g}) gives a  higher weight to long-living modes compared to fast decaying superradiant states \cite{PhD}. The main advantage in using a quantity like $g$ in (\ref{g}) is that it depends only on the eigenvalue spectrum of 
$H_{eff}$, i.e. it does not require the knowledge of the eigenfunctions of $H_{eff}$, which are far more difficult to obtain.

Figure 10 shows the behavior of $g$ as a function of system size for the scalar case (top) and the vectorial case (bottom) when cooperative pairs are excluded. In the scalar case, $g$  increases as a power law of the system size for dilute gases, but decays exponentially with the system size for dense clouds. In the vectorial case, $g$ varies as a power law of the system size  for both dilute and dense gases. It should be emphasized  that in the latter case the  resonance overlap does not decrease when the sample size is increased, even for the densest samples investigated. The results are similar when cooperative pairs are taken into account \cite{PhD}.

\begin{figure}[h!]
\centering
\begin{minipage}[b]{8.5cm}
\includegraphics[width=9.0cm]{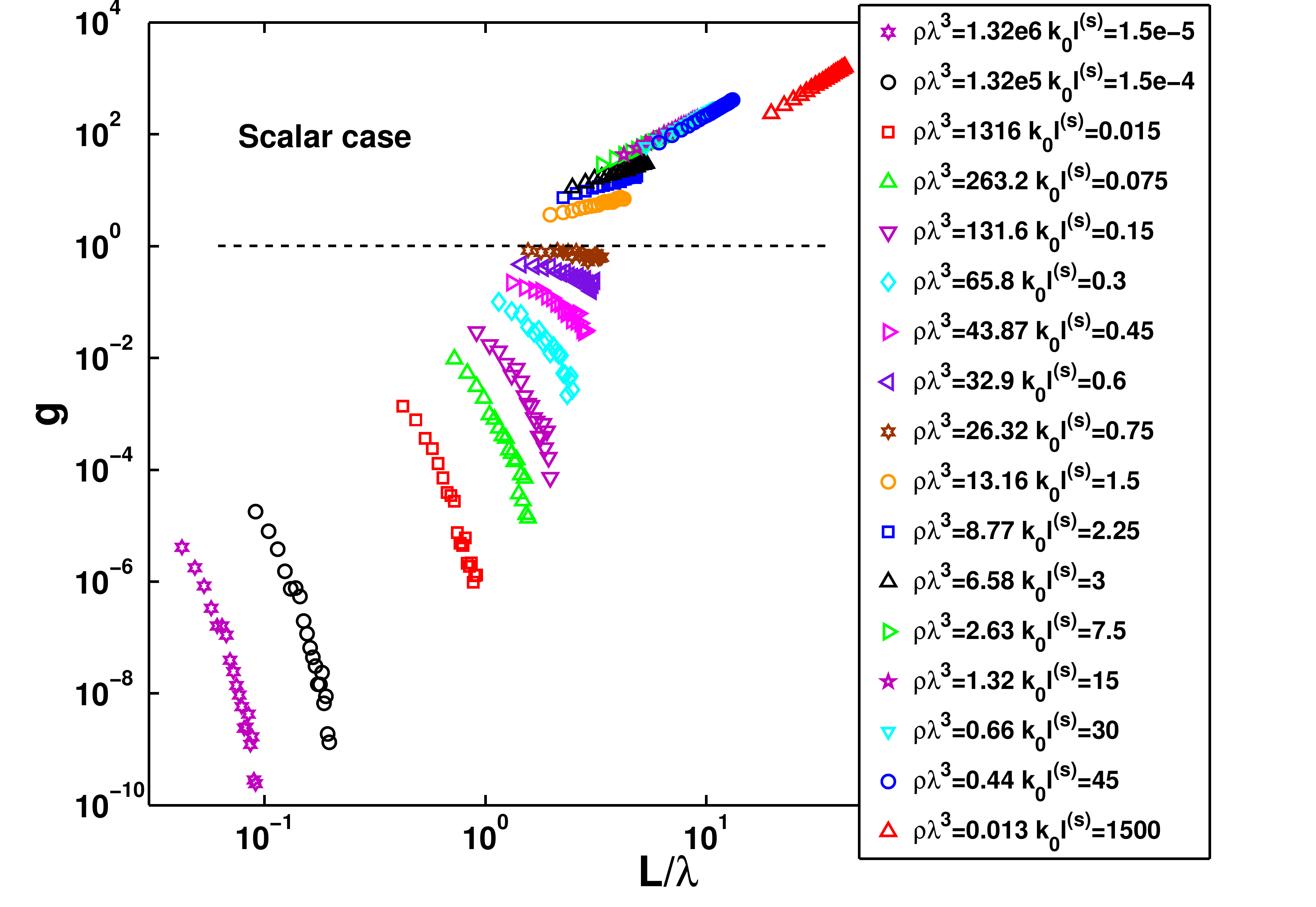}
\end{minipage}
\hspace{0.5cm}
\begin{minipage}[b]{8.5cm}
\includegraphics[width=9.0cm]{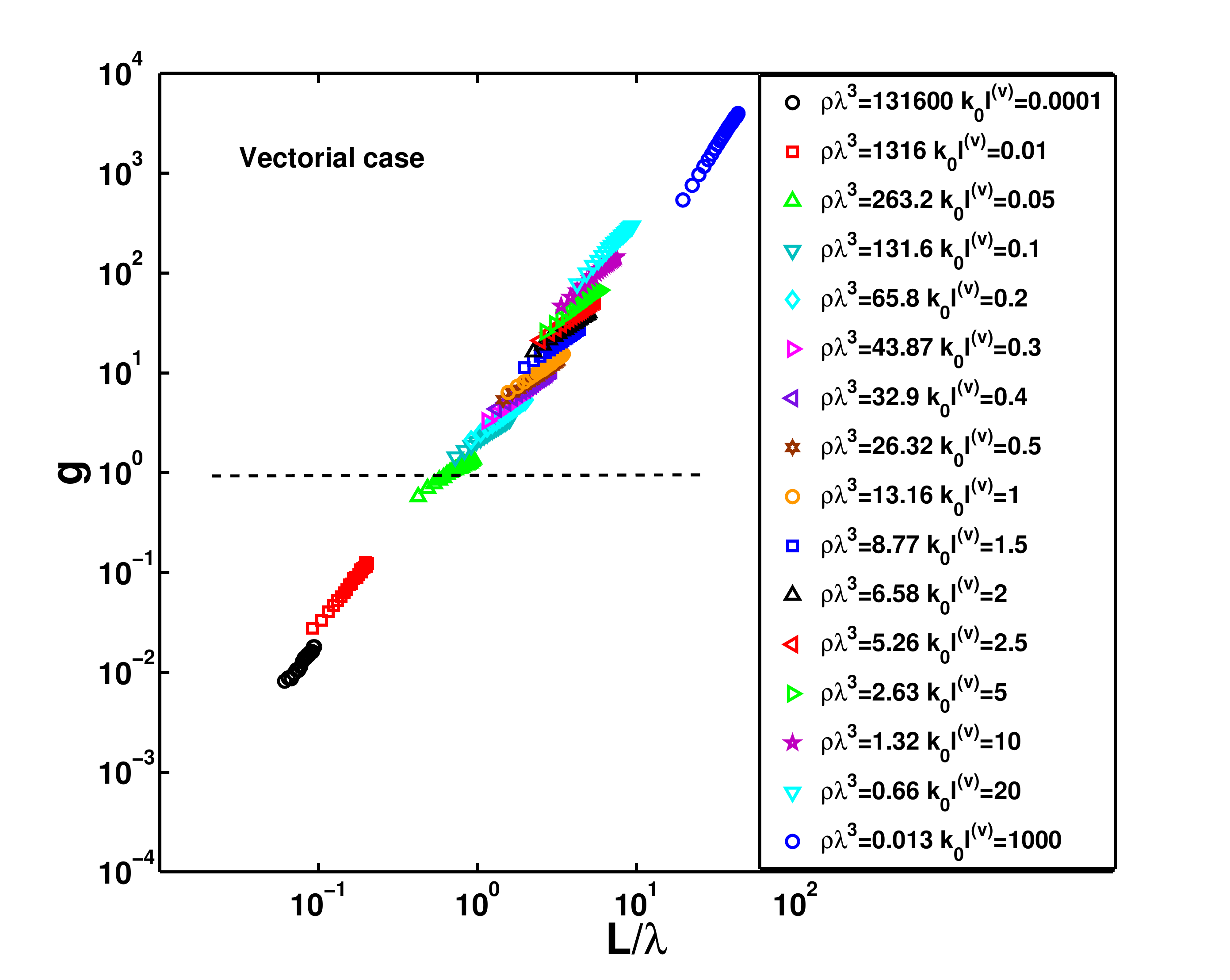} \caption{\em (Color
   online) Degree of resonance overlap, $g$, as a function of system size for the scalar case (top) and the vectorial case (bottom) when cooperative pairs are excluded.}
\label{fig10}
\end{minipage}
\end{figure}

Figure 11 shows  $g$ as a function of the Ioffe-Regel number  for the scalar case (top) and the vectorial case (bottom) when cooperative pairs are excluded. In the scalar case, the curves of $g$ corresponding to
different system sizes cross at $k_0l\sim 1$, as expected from the Ioffe-Regel criterion. In the vectorial case, however, 
no crossing point is observed.

\begin{figure}[h!]
\centering
\begin{minipage}[b]{8.5cm}
\includegraphics[width=8.5cm]{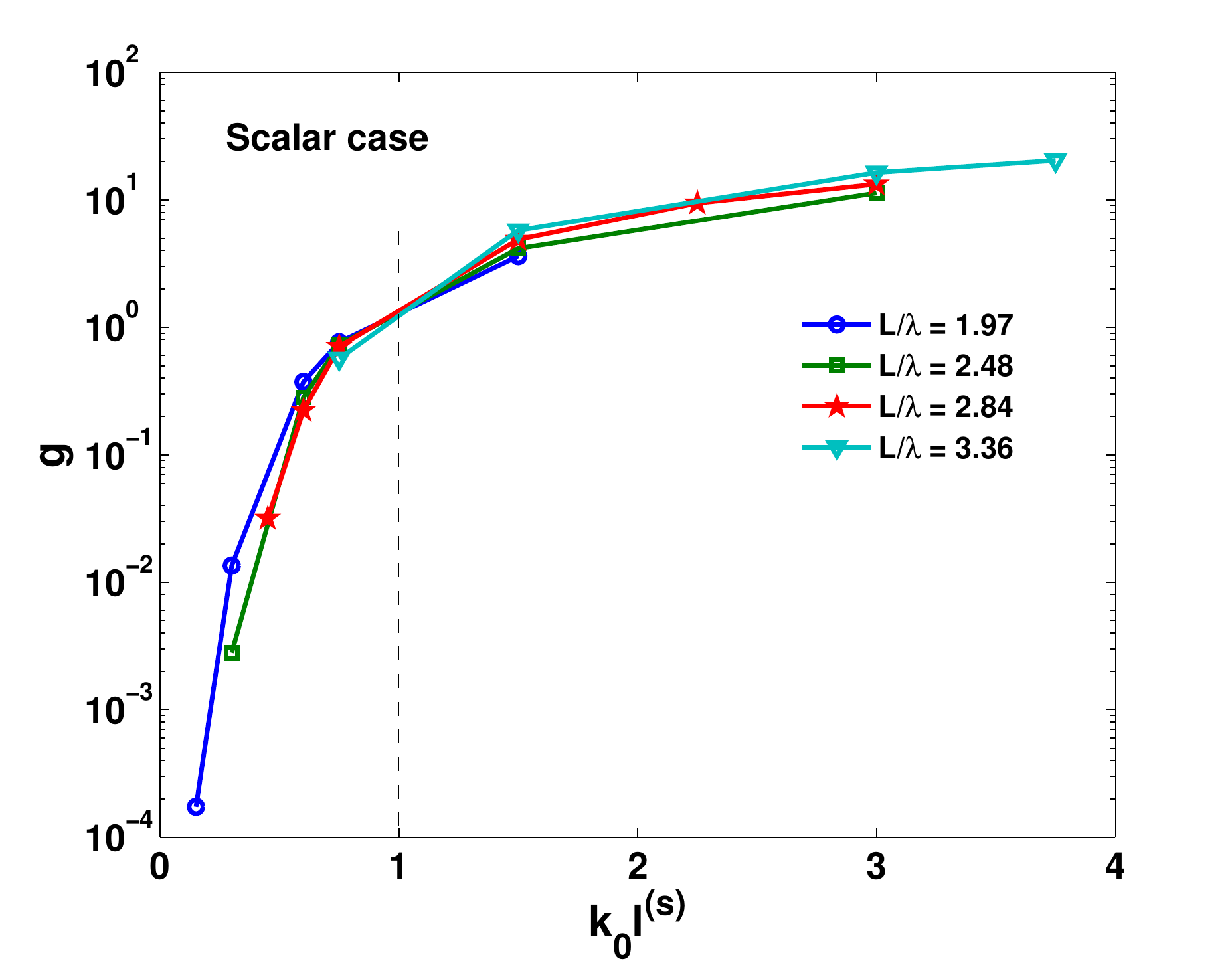}
\end{minipage}
\hspace{0.5cm}
\begin{minipage}[b]{8.5cm}
\includegraphics[width=8.5cm]{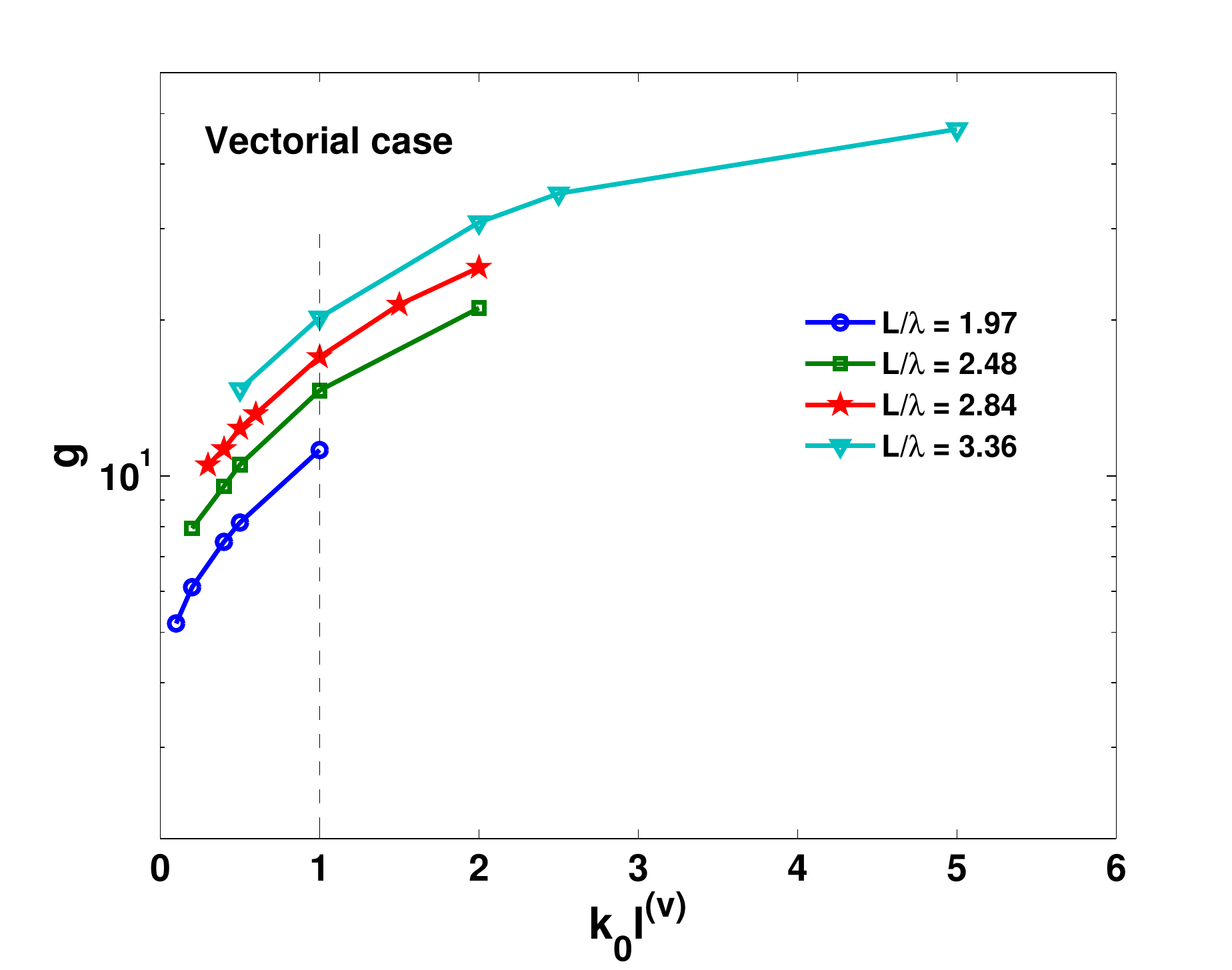} \caption{\em (Color
   online) The degree of resonance overlap, $g$, as a function of the Ioffe-Regel number for the scalar case (top) and the vectorial case (bottom) when cooperative pairs are excluded.}
\label{fig11}
\end{minipage}
\end{figure}

The clear scaling behavior observed for $g$ is rather unexpected and very interesting. It is first to be noted that it shows up over a broad range of system sizes, covering both the large system regime and the Dicke regime. It is interesting to analyze this scaling behavior using an analog of the Gell-Mann and Low function, $\beta(g)\equiv d \ln g/ d \ln L/ \lambda$, widely used in the theory of phase transitions \cite{gang}. We have extracted it from Fig. 10 and plotted this function in Fig. 12.

\begin{figure}[h!]
\centering
\begin{minipage}[b]{8.5cm}
\includegraphics[width=8.5cm]{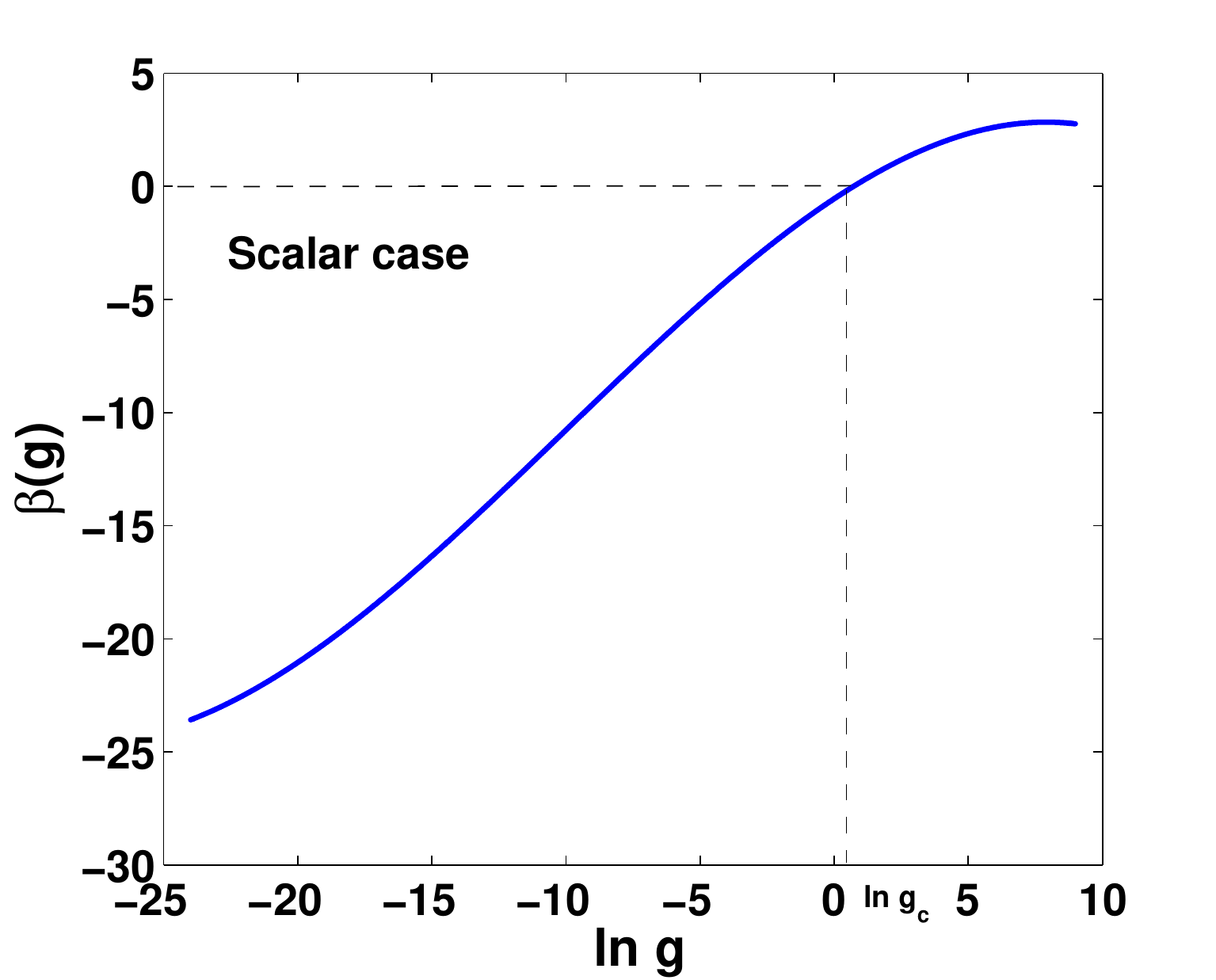}
\end{minipage}
\hspace{0.5cm}
\begin{minipage}[b]{8.5cm}
\includegraphics[width=8.5cm]{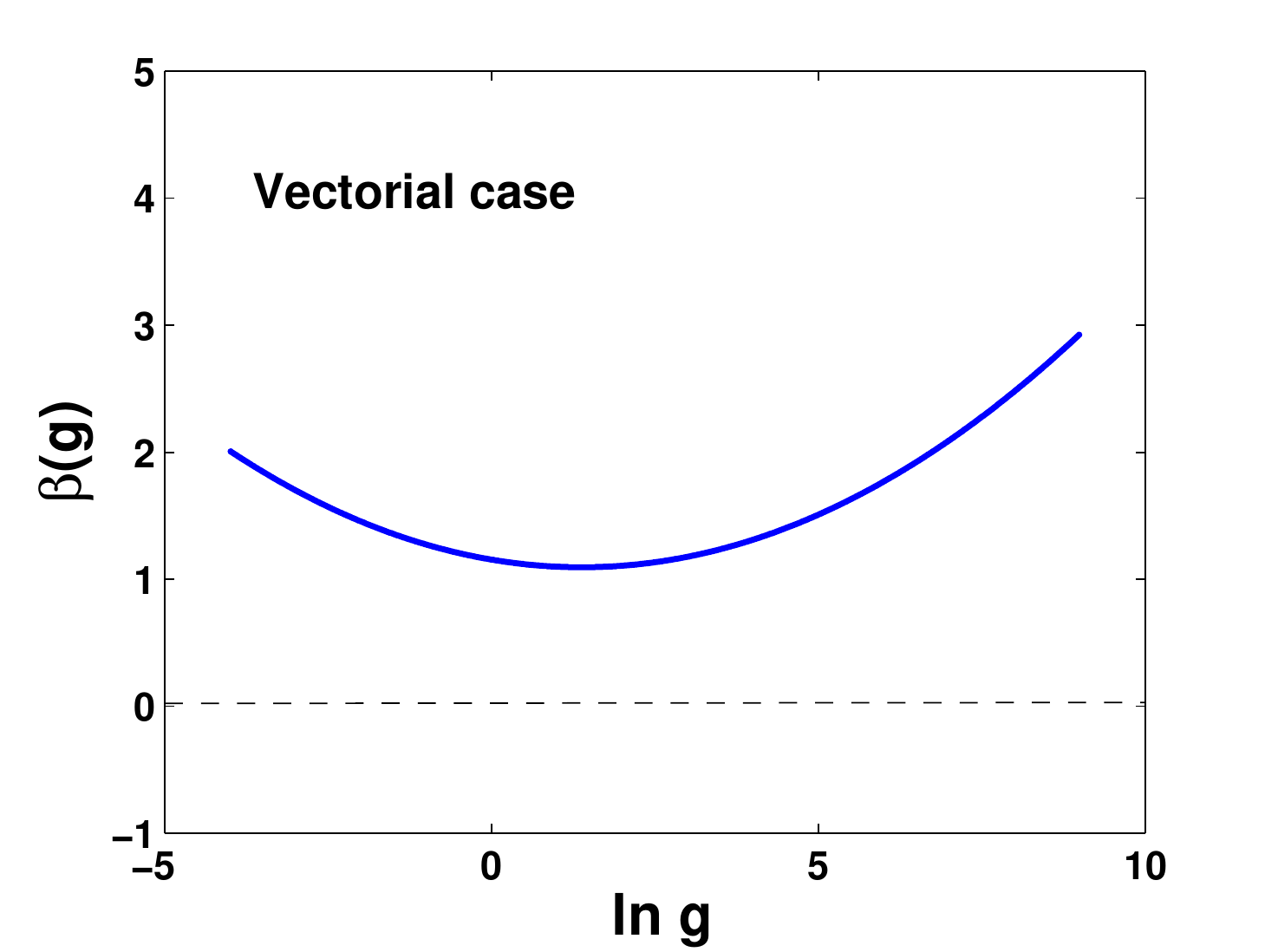} \caption{\em (Color
   online) $\beta(g)\equiv d \ln g/ d \ln L / \lambda$ as a function of $\ln g$ in the scalar case (top) and the vectorial case (bottom). Assuming that $\beta(g)$ is a continuous and monotonic function of $L$, in the scalar case there is a value  $g_c$ at which $\beta (g_c)=0$, indicating that  $g_c$  becomes independent of the system size $L$.}
\label{fig12}
\end{minipage}
\end{figure}

In the scalar case, $g$ increases as a power law of the system size $L$ in the limit of a dilute gas and it decreases exponentially with size for dense clouds. This change of behavior implies, assuming that $g$ is a continuous and monotonic function of $L$, that there exists a characteristic value $g_c$ at which $\beta (g_c)=0$, i.e., for which $g_c$ is independent of the system size $L$. Such a behavior is very reminiscent of Anderson-like phase transition driven by disorder. It should be noted, however, that the present case includes also the Dicke regime where cooperative effects play a major role. In that sense, the present case differs essentially from an Anderson, disorder-driven, phase transition. In contrast,  in the vectorial case, $\beta(g)$ is always positive. 

A similar analysis has been recently  presented \cite{Skipetrov2014}. The authors of  \cite{Skipetrov2014} have shown numerically that  localization of light can be achieved in a random three-dimensional atomic ensemble only for a scalar radiation field; it cannot be achieved when the vectorial properties of the electromagnetic wave are taken into account. The results presented in this paper conform to the results in  \cite{Skipetrov2014}, although we do not observe the second crossing point at high densities reported in  \cite{Skipetrov2014} for the scalar case, a fact we associate to the different ways of selecting the modes considered  for defining the resonance overlap criterion. 

Using the behavior of $g$ close to $g_c$ in the scalar case, it would be possible in principle to extract some more information regarding the observed critical behavior, e.g., the singular behavior of the localization length and the corresponding critical exponents. This would require more refined numerics not yet available.

\section{Discussion}

In this paper we have studied numerically the spectrum of the effective atomic Hamiltonian $H_{eff}$ given in (\ref{eqc13}) that describes the dipolar interaction of a gas of $N\gg 1$ atoms with the radiation field, both in the scalar and vectorial cases.

We have found that for dense gases, the resonance width distribution follows, both in the scalar and vectorial cases,  a power law  $P(\Gamma) \sim \Gamma^{-4/3}$. This power law  is different from the known  $ P(\Gamma) \sim \Gamma^{-1}$  distribution, which is interpreted  as a signature of Anderson localization of light in random systems \cite{Orlowski}. Even though this result is not energy specific, it suggests that long-living collective states of excitations are dominated by cooperative effects rather than disorder. As this power law holds for the case where cooperative pairs are excluded, it is related to cooperative effects between  more than two atoms.

We have also shown that the center of the energy distribution in dilute gases is described by Wigner's semicircle law not only in the scalar case, as suggested in \cite{skipetrov}, but in the vectorial case as well.  For dense clouds, we have shown that Wigner's semicircle law is replaced in the vectorial case by the Laplace distribution. Since in all cases, $P(E)$ is determined solely by the optical thickness, the energy distribution results mainly from cooperative effects.

Finally, we  have shown that in the scalar case the degree of resonance overlap behaves as a power law of the system size for dilute gases, but decays exponentially with the system size for dense clouds. In the vectorial case $g$ varies as a power law of the system size for both dilute and dense gases. 
As these findings hold also for the Dicke regime (i.e., in a system size much smaller than the wavelength), where cooperative effects are dominant,
a full interpretation based only on a disorder-driven phase transition (e.g,. Anderson localization) \cite{gang} appears to be incomplete.


 Further research on disorder-driven phase transition as expected on the basis
of Anderson localization might focus on transport properties of  light through atomic clouds \cite{Courteille2010} or 
consider the possibility to combine additional diagonal disorder to the long-range dipole-dipole coupling, with the possibility of hybrid states, sharing properties of disorder and synchronization \cite{HybridStates}.
Further insight to the different roles played by disorder and cooperative effect could be obtained  by  exact diagonalization of the effective Hamiltonian. Such calculations have already been done for the scalar case in the limit of dilute gases \cite{skipetrov, skipetrovb}. The vectorial case, however, still poses a substantial challenge.



\ \
\acknowledgments
 This work was supported by Israel Science Foundation Grant No. 1232/13.

\end{document}